\documentclass[%
 reprint,   
superscriptaddress,
 amsmath,amssymb,
 aps,
floatfix,
twocolumn,
]{revtex4-1}

\usepackage{graphicx}
\usepackage{dcolumn}
\usepackage{bm}
\usepackage{hyperref}
\usepackage[mathlines]{lineno}
\usepackage{color} 
\usepackage{tikz}
\usepackage{url}
\usepackage{amsmath}
\usepackage{amssymb}
\usepackage{enumerate}
\usepackage{tabularx}

\newcommand{\ava}{\langle a \rangle}
\newcommand{\asq}{\langle a^2 \rangle}

\begin{document}

\preprint{APS/123-QED}

\title{Simplicial Activity Driven Model}

\author{Giovanni Petri}
\email{giovanni.petri@isi.it}
\affiliation{ISI Foundation, Turin, Italy}
\author{Alain Barrat}%
\affiliation{Aix Marseille Univ, Universit\'e de Toulon, CNRS, CPT, Marseille, France}
\affiliation{ISI Foundation, Turin, Italy}

\date{\today}

\begin{abstract}

Many complex systems find a convenient representation in terms of networks: structures made by pairwise
interactions (links) of elements (nodes). For many biological and social systems, elementary 
interactions involve however more than two elements, and simplicial complexes 
are more adequate to describe such phenomena. Moreover, these interactions often change over time. 
Here, we propose a framework to model such an evolution: the Simplicial Activity Driven (SAD) model,  
in which the building block is a simplex of nodes representing a multi-agent interaction.
We show analytically and numerically that the use of simplicial structures leads to crucial structural differences  
with respect to the activity-driven (AD) model, a paradigmatic temporal network model involving only 
binary interactions. It also impacts the outcome of paradigmatic processes modelling disease propagation or social contagion.
In particular, fluctuations in the number of nodes involved in the interactions can affect the outcome of models of
simple contagion processes, contrarily to what happens in the AD model.


\end{abstract}

\maketitle


The use of a network representation has become commonplace for 
describing and studying a large number 
of complex systems: the elements of the systems are seen as 
nodes, and links
between nodes represent pairwise interactions \cite{Albert:2002,BBV}.
However, in many contexts, representing interactions as pairwise 
does not tell the whole story.
Examples include collaborations among groups of actors in 
movies \cite{Ramasco:2004}, spiking neuron populations \cite{giusti2015clique,reimann2017cliques} and 
co-authorships in scientific publications \cite{patania2017shape}. 

Let us consider the latter for illustration purposes: 
In a network representation, a paper co-authored by $n$ scientists yields a clique of  $n(n-1)/2$ links, which is however 
treated in the same way as $n(n-1)/2$ papers authored by pairs of scientists (or any other
combination of subgroups among these $n$ scientists leading to the same number of links). 
While this is equivalent for $n=2$, the number of co-authors of a scientific paper 
is often larger than $2$. For instance, data (see Supplemental Material - SM) show that
the average number of authors of APS papers has steadily increased 
from $2$ to $6$ between the 1940s and now.
In such cases ($n > 2$), simplicial representations are more apt to preserve the information observed in data. 
To take this issue into account, simplicial descriptions have recently been adopted 
in models of emerging geometry \cite{bianconi2016network,bianconi2015complex}, 
null models for higher order interactions \cite{Courtney:2016,young2017construction}, network inference \cite{benson2018simplicial}, brain structure and dynamics \cite{lord2016insights,giusti2015clique,sizemore2018cliques}.

We recall that formally, a $(d-1)$-dimensional simplex $\sigma$ is 
defined as the set of $d$ vertices $\sigma = [x_0, x_1, \dots, x_{d-1} ]$. 
A collection of simplices is a simplicial complex $K$ if for each simplex $\sigma$ all its possible subfaces (defined as subsets of $\sigma$) are themselves contained 
in $K$ (see SM). 
In the case of group interactions, this requirement can be considered trivially satisfied, as each group interaction implies all the possible sub-interactions. 
Finally, the $1$-skeleton of a simplicial complex is the collection of all its edges, i.e., the underlying network.

Networks and interactions moreover evolve in time, and the field of temporal networks has indeed 
become very active \cite{Holme:2012,Holme:2015}. In particular,
several models of time-evolving networks have been put forward, based on microscopic rules for the establishment and end of interactions between pairs of nodes
\cite{Stehle:2010,Perra:2012,Vestergaard:2014}. 
Among these, the activity driven (AD) temporal network model \cite{Perra:2012} has attracted a lot of attention. In this model, each agent (node) is assigned an activity potential that determines at each time its probability to create pairwise interactions with other agents selected at random. The AD model and its extensions \cite{Karsai:2014,Sun:2015,Nadini:2018,Kim:2018} have 
become a paradigm of temporal networks and have been 
used to study the impact of the network's temporal evolution
on dynamical processes occurring on top of it \cite{Perra:2012,Perra:2012b}. 

Models of temporally evolving simplicial
interactions are however still missing. 
Here we bridge this gap by proposing a modeling
framework for temporal group activation data: the Simplicial Activity Driven model. Our aim is to provide a simple 
framework that can be used as a basis for richer temporal models taking into account the simplicial nature of interactions, and on which dynamical processes can be studied by analytical and numerical means to shed light on the impact of both simplicial and temporally evolving interactions.

\begin{figure}[thb]
\centering
\includegraphics[width=0.90\columnwidth]{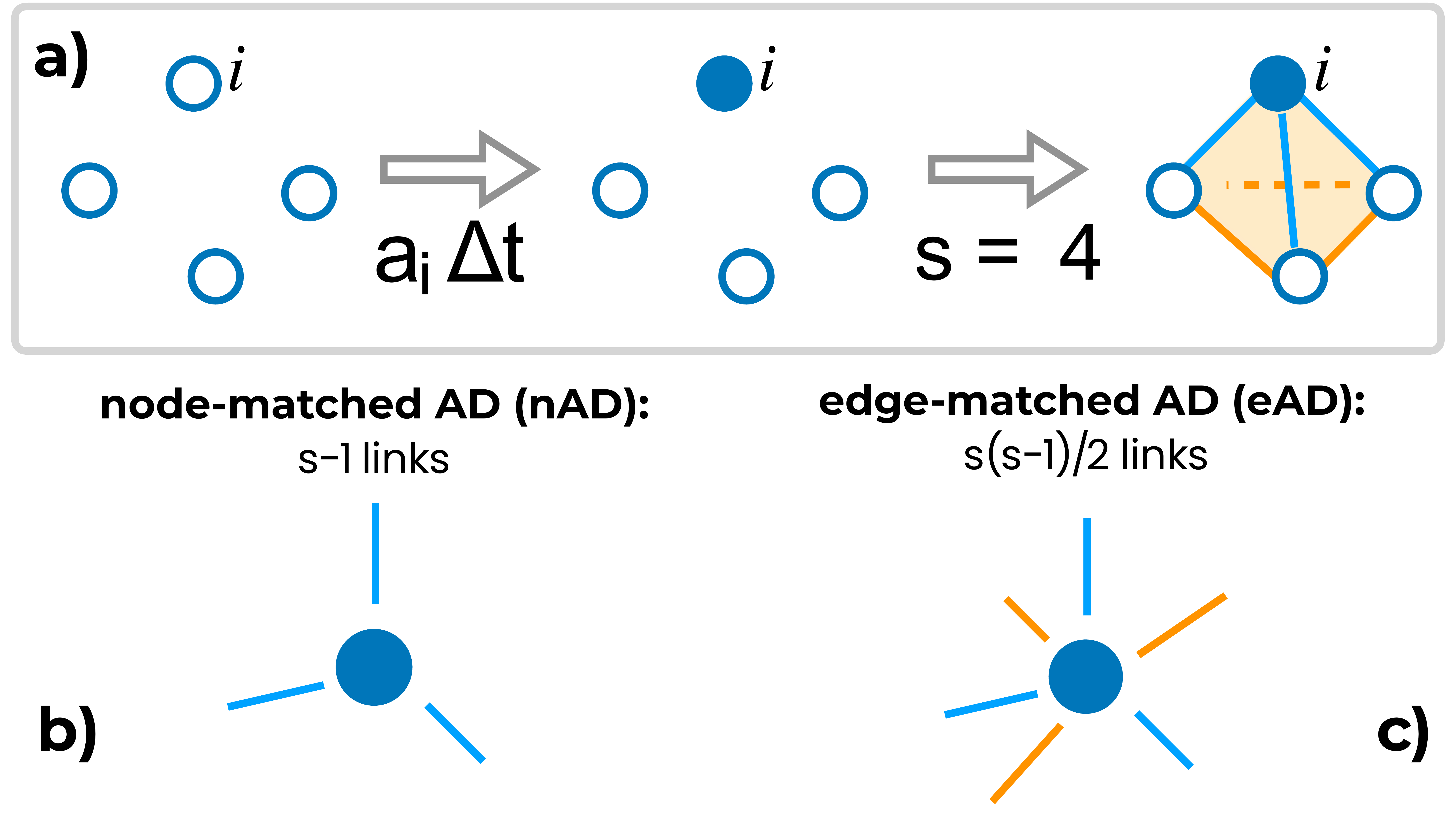}
\caption{\textbf{SAD model.} a) At each timestep, a node $i$ activates with probability $a_i\Delta t$. Upon activation it creates a coherent unit of $s$ nodes (an $(s-1)$-simplex), with links between all pairs. b) In contrast, in the standard AD model (nAD) only the $s-1$ edges stemming from the activated node are added. 
c) In the eAD model 
$\binom{s}{2}$ links stem from the activated node, conserving at each interaction the number of links of the SAD model.
}
\label{fig:sad-cartoon}
\end{figure}

In its simplest version, the model considers $N$ nodes, whose interactions change over time
as follows:
\begin{enumerate}[(i)]
\item Each node $i$ is endowed with an activity rate $a_i$ taken from a predefined distribution $F$;

\item At each time step $\Delta t$, each node $i$ fires with probability $a_i \Delta t$; when it fires, 
it creates an $(s-1)$-simplex (in networks' terms, 
a clique of size $s$) with $s-1$ other nodes chosen uniformly at random. Each activation hence contributes $s(s-1)/2$ interactions to the network;

\item At the following time step, the existing simplices 
are erased and the process starts anew.

\end{enumerate}
In the framework of collaborations, nodes can represent scientists and the activity $a_i$ their propensity to create collaborations: step $(ii)$ corresponds to the 
creation of a collaboration of $s$ scientists resulting in the co-authorship of a paper. 
We underline the main difference with the AD model in Fig. \ref{fig:sad-cartoon}: 
in the AD model, each active node creates a set of {\em binary} interactions with the chosen nodes
(in the language of collaborations, $s-1$ papers  with each $2$ authors) while, in the SAD model, nodes 
that are not active but are targeted by an active node obtain 
links to all the other nodes in the simplex, creating a coherent activation unit. 
The parameter $s$ defines the size of the collaborations
and can either be fixed or a random variable extracted at each activation from a distribution $p(s)$.
As for the AD model, the SAD model 
is Markovian: agents do not have memory of the previous time steps, and it
can be refined by adding memory or community effects \cite{Karsai:2014,Kim:2018}.

In the following, we first study this model from a structural point of view,highlighting also the differences
between considering the obtained system as a network (given by the 1-skeleton of the simplicial complex) and taking into account its simplicial nature.
We also provide an analysis of a paradigmatic 
dynamical process occurring on top of the SAD model. 
Since the AD model has been widely studied in the literature as a paradigm for temporally evolving networks, we underline in each case how
the introduction of coherent units of $s$ nodes as building blocks, instead of sets of binary interactions, yields
radically different structural properties and impacts the properties of dynamical processes. 

The comparison with the activity-driven model can be done in two ways: we can
indeed consider AD models designed to involve either the same
number of nodes (node-matched AD model, nAD) or the same number of interactions (edge-matched AD model, eAD) 
as the SAD model at each time step (see Fig. \ref{fig:sad-cartoon}).
In the former, for each activation with group size $s$ 
in the SAD model, we consider an AD activation with size 
$m =s-1$, i.e., the activated node creates interactions
with $s-1$ other nodes chosen at random: this leads to the 
same total number of contacted nodes
per activation in the nAD and in the SAD model. 
In the eAD, for each activation with group size $s$, we consider 
instead an AD activation with $m=\binom{s}{2}$, hence preserving the total number of interactions of each activation. 

\paragraph*{Structure}
Let us focus on the structural properties of the SAD 
model, once aggregated over a fixed number $T$ of timesteps.
We first consider the $1$-skeleton of the SAD model, i.e., the network obtained as the union of all its edges, 
in order to compare its properties to AD networks, and then consider pure
simplicial properties -- not reducible to a network approach.

In the aggregated SAD, each node $i$ is linked
by an edge to all the nodes with whom it has interacted
at least once during the aggregation time-window. The
degree of $i$ in the corresponding aggregated network corresponds
to the number of distinct nodes with whom $i$ has 
interacted; in the interpretation of a scientific collaboration network, it gives the total number of distinct collaborators of a scientist. 

\begin{figure}[tb]
\centering
\includegraphics[width=0.9\columnwidth]{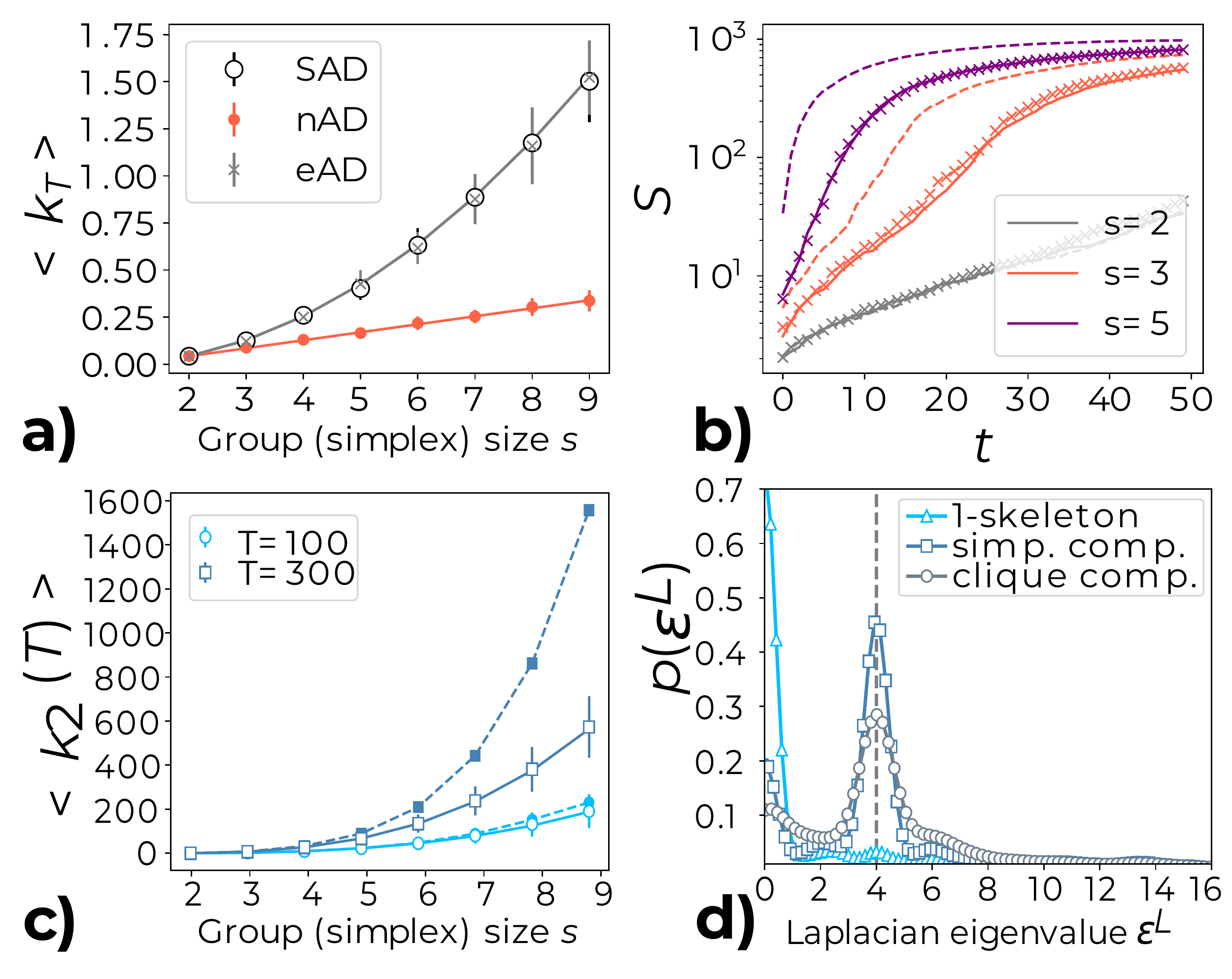}
\caption{\textbf{Structural properties of SAD model.}
In all plots we use $N=2000$ nodes and activities sampled from $F(a) = (a/a_0)^{-\alpha}$, $\alpha=2.1$ and $a_0=5\cdot 10^{-3}$.
(a) Average aggregated degree $\langle k_T\rangle$ 
for the SAD model and corresponding nAD and eAD models vs. simplex size $s$, with $T=10$.
Symbols: numerical values; lines: theoretical predictions (for the SAD, Eq. \ref{eq:integrated-degree} averaged over all nodes).
(b) Temporal growth of the aggregated GCC size $S$ for the SAD (solid lines), nAD (crosses) and eAD (dashed lines) for various fixed simplex size $s$. 
(c) Empty symbols: average $\langle k_2(i,T)\rangle$ over all nodes $i$ of the number of $2$-simplices to which $i$ belongs, 
in the SAD aggregated until time $T$, for various
values of $s$ and of $T$. Continuous lines: prediction \eqref{eq:k2} averaged over nodes. The filled symbols give instead
the average number of triangles to which a node $i$ belongs in the $1$-skeleton of the aggregated SAD.
(d) Eigenspectrum of the simplicial Laplacian $\mathcal{L}_1$ computed on the aggregated SAD $1$-skeleton (here with $s=4$), on the actual aggregated SAD simplicial complex and on the clique complex of the $1$-skeleton  of the aggregated SAD simplicial complex (in which each $(k+1)$-clique of the $1$-skeleton is promoted to a $k$-simplex). Aggregation time: $T=100$. 
}
\label{fig:aggregated-network}
\end{figure}

Denoting by $k_T(i)$ the expected aggregated degree at time $T$
of node $i$ with activity $a_i$, we compute it  
by separating it into two contributions.
The first comes from node $i$'s own activation events, 
which occur at each timestep $\Delta t$ with probability
$a_i \Delta t$: 
after $T$ timesteps, 
$i$ will have activated $\sim T a_i$ times; for fixed simplex size $s$, this means it will have made $T a_i \bar{m}$ interactions ($\bar{m}=s-1$). 
The second contribution comes from the activations of other nodes: 
every node $j\neq i$ will have activated $T a_j$ 
times; in each activation of $j \neq i$, $i$ was selected with probability $\bar{m}/(N-1)$ and, 
if selected, provided with $\bar{m}$ interactions. 
Hence, at time $T$ node $i$ will have accumulated $\kappa_T(i)$ interactions with:
\begin{eqnarray}
\kappa_T(i) = \bar{m}a_i T + \sum_{j\neq i} \frac{\bar{m}^2 T a_j}{N-1}
 \simeq \bar{m}T(a_i + \bar{m} \langle a \rangle)
 \end{eqnarray}
where the approximation holds for $N\gg 1$. For any 
node distinct from $i$, 
the probability not to have been involved in any of 
these interactions is 
$(1 - 1/(N-1))^{\kappa_T(i)}$, and 
hence finally the number of distinct
nodes having interacted with $i$ is 
\begin{eqnarray}
k_T^{SAD}(i) & = &  (N-1) 
\left[ 1 - \left(1 - \frac{1}{N-1}\right)^{\kappa_T(i)} \right] \\
 \label{eq:integrated-degree}
 &\simeq & N \left[ 1 -  e^{- \frac{T\bar{m} (a_i + \langle a \rangle \bar{m})}{N}} \right] 
\end{eqnarray} 
where the approximation holds for large $N$ and small $T/N$ 
(in the SM we also give the derivation for the aggregated degree distribution).
%
In Fig. \ref{fig:aggregated-network}a we show the excellent agreement between the prediction for the aggregated degree averaged over all nodes,
$\langle k^{SAD}_T(i)\rangle$, at fixed $s$,  and numerical simulations. 
We also compare it with the nAD and eAD models, for which
$k_T^{n/eAD}(i) = N(1 - e^{-T m (a_i +\langle a\rangle)/N})$
\cite{Perra:2012} with $m=\bar{m}$ for nAD and 
$m=\binom{s}{2}$ for eAD. 

Interestingly, $k_T^{SAD}$ depends on $\bar{m}^2$: thus, if the simplex size $s$ is allowed to fluctuate, 
the size of such fluctuations will affect the aggregated degree (see SM).
This phenomenology is in contrast with the
nAD model, which has no dependence on the second moment of $s$, 
while the eAD inherits it from the matching of
the number of edges created at each
activation (since each activation creates $m=s(s-1)/2$ edges,
the resulting total number of activations and the integrated degree
depend on the fluctuations of $s$).

From this aggregated network degree point of view, we thus observe 
a similar behaviour for the SAD and eAD models. Figure
\ref{fig:aggregated-network}b however highlights that this is
not the whole story, and that the SAD model building mechanism
leads to an important structural difference with the eAD model:
as each activation creates
$\binom{s}{2}$ interactions that involve only $s-1$ nodes, the size
of the largest connected component (GCC) in the SAD
integrated until $T$ grows with $T$ much more 
slowly than in the eAD model, for which each activation 
creates a star reaching $\binom{s}{2}$ nodes; in fact, it grows
in the same way as in the nAD,
despite creating more interactions at each step (for $s >2$, as for $s=2$ the 
three models are the same). Overall, the structural properties of the SAD model, from the point of view of its $1$-skeleton, present thus
both similarities and important differences with AD models with the same numbers of events.

Let us now focus on purely simplicial structural properties of the SAD model.
First, we compute the average number $k_2(i,T)$ of $2$-simplices to which a node $i$ belongs
in the SAD aggregated until $T$, in a way similar to the computation
of $k_T(i)$. We obtain (see SM for details):
\begin{equation}
k_2(i,T) = \binom{N-1}{2} \left( 
1 - e^{-  \frac{(s-1)(s-2)}{(N-1)(N-2)} T (a_i + \bar{m} \langle a \rangle )   }  \right) .
\label{eq:k2}
\end{equation}
It is important to note that $k_2(i,T)$ corresponds to the number of distincts cliques of three nodes to which $i$ has participated from 
time $0$ to $T$, which is different from the number
of triangles to which $i$ belongs in the $1$-skeleton  of the aggregated SAD: indeed, a triangle $(i,j,k)$ can be obtained even if links $(i,j)$, $(i,k)$ and $(j,k)$
are never present in the same $(s-1)$-simplex.
We show in Fig. \ref{fig:aggregated-network}c that the average of $k_2(i,T)$ over all nodes is correctly predicted by Eq. \eqref{eq:k2}
for various values of $T$ and of $s$ (see SM for a more extensive validation of Eq. \eqref{eq:k2}). The figure also shows that
the average number of triangles to which a node belongs grows faster with both $s$ and $T$ than $\langle k_2(i,T)\rangle$, highlighting the differences
between simplices and triangles and thus the importance of the simplicial nature of the SAD model.

We moreover present in Fig. \ref{fig:aggregated-network}d the eigenspectrum of the simplical Laplacian (see SM) of the aggregated SAD, a cornerstone of studies of
how dynamical processes are affected by the underlying structure. This figure highlights how the eigenspectrum differs, depending on whether 
we compute the Laplacian on the $1$-skeleton,  on the aggregated SAD simplicial complex, or on the
clique complex obtained by considering each clique of the $1$-skeleton as a simplex. These differences illustrate further how the SAD contains
information not reducible to its $1$-skeleton, i.e., to a network.

 To further our analysis, we now explore
how dynamical processes are impacted by the SAD model.

\paragraph*{Dynamical processes}
We consider  the paradigmatic 
susceptible-infected-susceptible (SIS) model 
for disease spreading \cite{barrat2008dynamical}. 
In this model, nodes can be either 
susceptible (S) or infectious (I). Infectious individuals
propagate the disease to susceptible ones at rate $\beta$
whenever they are interacting, and recover spontaneously
at rate $\mu$, becoming again susceptible. 
In a homogeneous population, the epidemic
threshold $\lambda_c$ separates an epidemic-free state at low 
values of the parameter $\lambda = \beta/\mu$ from an endemic state
at high values of $\lambda$. 

To calculate the SIS epidemic threshold for the SAD model, we
use a temporal heterogeneous mean-field approach similar to 
the one used for the AD model \cite{Perra:2012}: 
nodes are classified according to their activity, and we 
denote by
$N_a$ the number of nodes with activity $a$; $I_a^t$ and
$S_a^t$ denote respectively the
numbers of infectious and susceptible nodes with activity $a$ at time $t$.
We thus have $N_a = S_a^t + I_a^t$, 
$N = \int da N_a$ is the total population, and 
$I^t = \int da I_a^t$ the total number of infectious at time $t$.

Let us consider for simplicity the case of the SAD model with fixed
clique size $s$. 
The variation during a time step $\Delta t$ 
of the number of infectious is given by the following equation 
taking into account the evolution of both interactions and spreading process:
\begin{eqnarray}
& &I_a^{t+\Delta t} - I_a^t  = - \mu \Delta t I_a^t  \nonumber
+\beta \Delta t S_a a (s-1) \int da' \frac{I_{a'}^t}{N} \\
&+& \beta \Delta t S_a \int da' a' \frac{I_{a'}^t}{N} (s-1)  \nonumber \\
&+& \beta \Delta t S_a \int da' a' \frac{S_{a'}}{N}(s-1) \int da'' \frac{I_{a''}}{N} (s-2) .
\label{SIS_a}
\end{eqnarray}
The first term corresponds to the recovery of
nodes with activity $a$. The second term corresponds to 
susceptible nodes with activity $a$ that become active at time $t$
(with probability $a \Delta t$)
and create a simplex of size $s$ (with $s-1$ other nodes) that
includes infectious nodes with any activity (hence
the integration over $a'$). 
The third term stems from the fact that 
susceptible nodes with activity $a$ can be chosen as clique partners
by infectious nodes with activity $a'$ that become active
(with probability $a'\Delta t$). While these three terms
also appear in the case of a spreading process on an AD network,
the last term is specific to the SAD model: it
describes the cases in which 
susceptible nodes with activity $a$ are chosen by 
a susceptible with activity $a'$, which becomes
active (with probability $a'\Delta t$) and creates
a simplex that also includes an infectious node with activity $a''$.

Straightforward computations
detailed in the SM yield then the epidemic 
threshold condition
\begin{equation}
\frac{\beta}{\mu}  >  \frac{2}{ s (s-1) \ava   + 
(s-1) \sqrt{s^2 \ava^2 + 4 (\asq - \ava^2)} } ,
\label{epi_sad_1}
\end{equation}
to compare with the result 
$\frac{\beta}{\mu}  >  1 /( m \ava   + m \sqrt{  \asq } )$
for an AD model with parameter $m$.

\begin{figure}[tb]
\includegraphics[width=0.9\columnwidth]{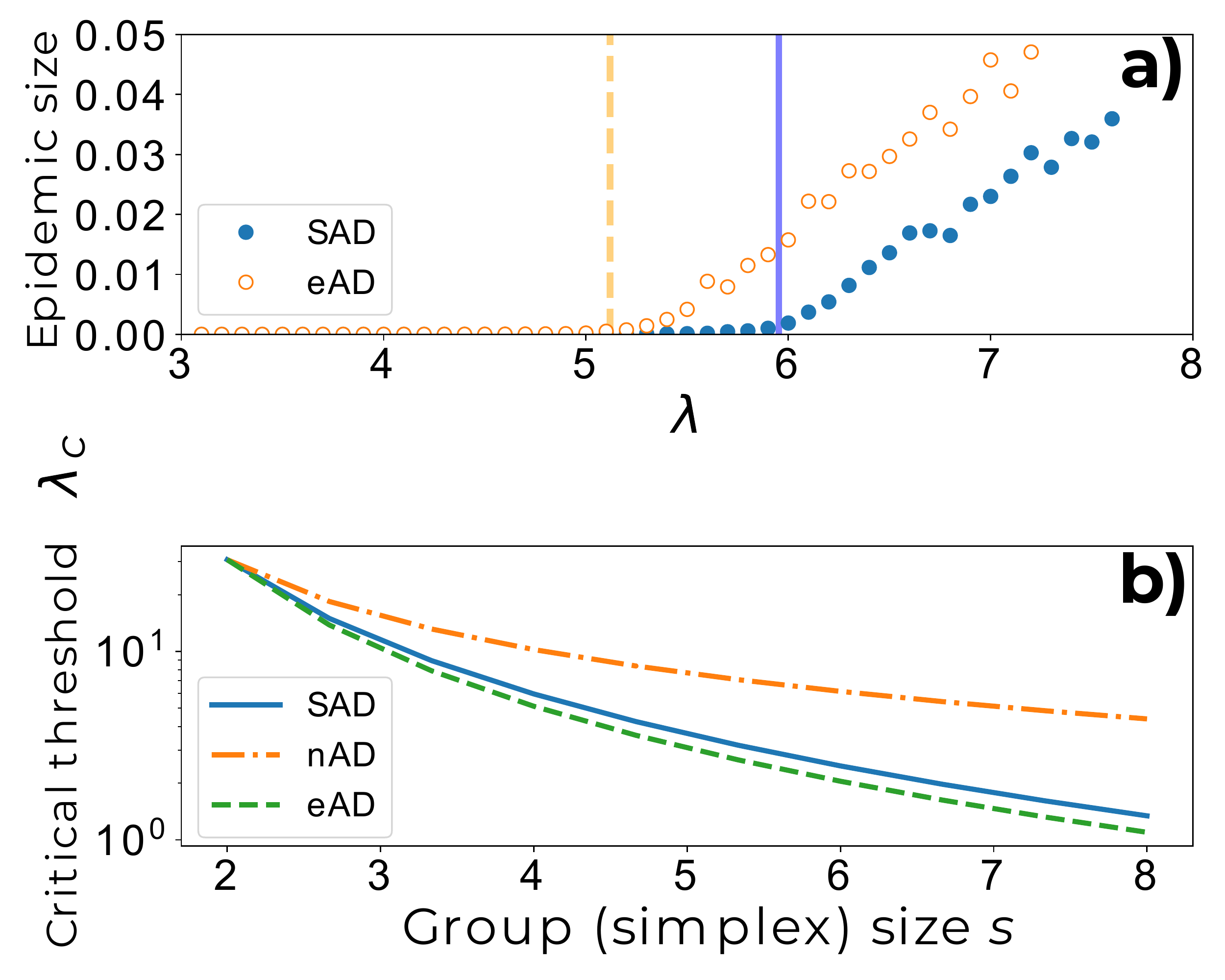}
\caption{\textbf{SIS epidemic threshold for AD and SAD models.} 
a) epidemic prevalence versus $\lambda=\beta/\mu$: 
the epidemic transition in the SAD model 
is delayed as compared with an SIS model on the corresponding eAD
model (here $N=1000$, $s=4$, $T=20000$).
Vertical lines correspond to the 
theoretical values of the epidemic thresholds (Eq. \eqref{epi_sad_1} for the SAD model).
b) Increasing the average connectivity of the underlying 
network lowers the epidemic threshold in all models; 
for the $s$-regular SAD model $\lambda_c$ (Eq. \eqref{epi_sad_1}) is always 
larger than for the corresponding eAD model. 
In both
panels, node activities were sampled from $F(a) = (a/a_0)^{-\alpha}$, $\alpha=2.1$ and $a_0=5\cdot10^{-3}$. 
} 
\label{fig:epi-thr}
\end{figure}

If the sizes of the cliques formed in the
SAD model are extracted at random at each activation
from a distribution $p(s)$, the r.h.s. of Eq.~\eqref{SIS_a}
needs simply to be integrated as $\int ds p(s)$, if the size
$s$ is independent of the activity $a$. The epidemic 
threshold becomes
\begin{equation}
\frac{\beta}{\mu}  >  \frac{2}{ \langle s (s-1)\rangle \ava   +  
\sqrt{\Delta}} ,
\label{epi_sad_2}
\end{equation}
with $\Delta = \langle(s-1)(s-2) \rangle  \langle(s-1)(s+2) \rangle 
\ava^2 + 4 \langle s-1 \rangle^2 \asq$ (see details in the
SM). Notably, it depends not only
on the average clique size but also on the second moment
of $p(s)$, and vanishes as $1/\langle s^2 \rangle$ 
if the clique size fluctuations diverge (see SM).
This is in contrast with the case of the SIS model on 
the AD model, in which fluctuations of the numbers of links
created at each time step would not change the epidemic threshold
($m$ just being replaced by its average). 

Figure \ref{fig:epi-thr}a displays the result of  
numerical simulations of an SIS model on temporal eAD and SAD
networks, showing agreement with the theoretical values of the epidemic threshold. 
We moreover compare in panel (b) the epidemic threshold obtained in 
a SAD model with fixed clique size $s$ with 
the epidemic threshold obtained in the nAD and eAD models.
In the former case, the epidemic
threshold is smaller in the SAD model, which can be related
to the fact that the SAD model
has more interactions than the nAD network. 
In the latter case on the contrary,
the fact that the $s(s-1)/2$ interactions are created as cliques hampers the spread
on the SAD with respect to the eAD network, leading  to a higher epidemic threshold for the SAD case 
(see SM for examples).
.


To conclude, we have presented a new model for temporal networks, 
based on the fact that the fundamental building blocks of many social
networks are coherent units of several individuals interacting as a group,
rather than dyadic interactions. Our Simplicial Activity Driven model considers indeed agents who, when active, create simplices
with other agents, yielding a simplicial complex once aggregated. 
We have shown how this mechanism leads to fundamental differences with respect to a well-known model in which 
active agents create sets of dyadic interactions, the activity-driven model, and how the structural properties of the SAD model differ
from those of its $1$-skeleton, showing the necessity to take into account its simplical nature and not to reduce it to a network interpretation. 
These differences appear not only at the structural level but
have also strong consequences on how dynamical processes unfold on these networks, as we have illustrated on a paradigmatic epidemic model 
(we also show in the SM that a social contagion process on the SAD displays a rich phenomenology, and 
can become extremely slow in the SAD with respect to an AD model). \\
As noted in its definition, the SAD model is Markovian, as the AD model: it does not yield non-Poissonian nor bursty temporal patterns. However,
thanks to this simplicity, our model lends itself to analytical investigations of its structural properties and of contagion processes, which
has allowed us to highlight the need to correctly take into account the simplicial nature of interactions in models as well as the fluctuations of the numbers
of nodes involved in these interactions. Moreover, it can serve as 
a starting point for a number of refinements, such
as adding memory effects, node categories and interacting probabilities depending
on these categories, or correlations between activity of an agent and
size of the simplicial complex it creates. Moreover, it would be interesting
to study further dynamical processes on the SAD model and its variations.
Finally, the SAD model constitutes a first null model for the homology of temporal complex systems with high-order interactions. 
We hope that our work will stimulate research in such directions.


\clearpage
\newpage

\begin{widetext}

\setcounter{figure}{0}
\setcounter{table}{0}
\setcounter{equation}{0}

\makeatletter
\renewcommand{\thefigure}{S\@arabic\c@figure}
\makeatother

\makeatletter
\renewcommand{\thetable}{S\@arabic\c@table}
\makeatother

\makeatletter
\renewcommand{\theequation}{S\@arabic\c@equation}
\makeatother

\section*{Simplicial Activity Driven Model: Supplemental Material}

\section{Co-authorship data}

We consider a dataset containing all publications in 
American Physical Society  (APS) journals from year 1893 to 2016. 
The dataset is part of the ``APS Data Sets for Research'' and can be freely obtained from \url{https://journals.aps.org/datasets}. 
The dataset contains 587675 papers by 413513 different authors. 
Figure \ref{fig:APS} illustrates the evolution
of the distribution of the number of co-authors as deduced
from this dataset: over the course of a few decades, this distribution has evolved
from a narrow one in the 1940s to progressively broader distributions in 1960s and up present times, with both average and fluctuations
increasing with time.
The average number of authors of a paper
has steadily increased from $2$ to $6$ between the 1940s and now.
Moreover, the inset of Fig. \ref{fig:APS} shows 
the temporal evolution of the ratio $\beta_s = \frac{\langle s^2 \rangle}{\langle s \rangle}$ between the first two moments of $p(s)$, highlighting
the increase in the heterogeneity of the distribution.

\begin{figure}[hb]
\centering
\includegraphics[width=.5\columnwidth]{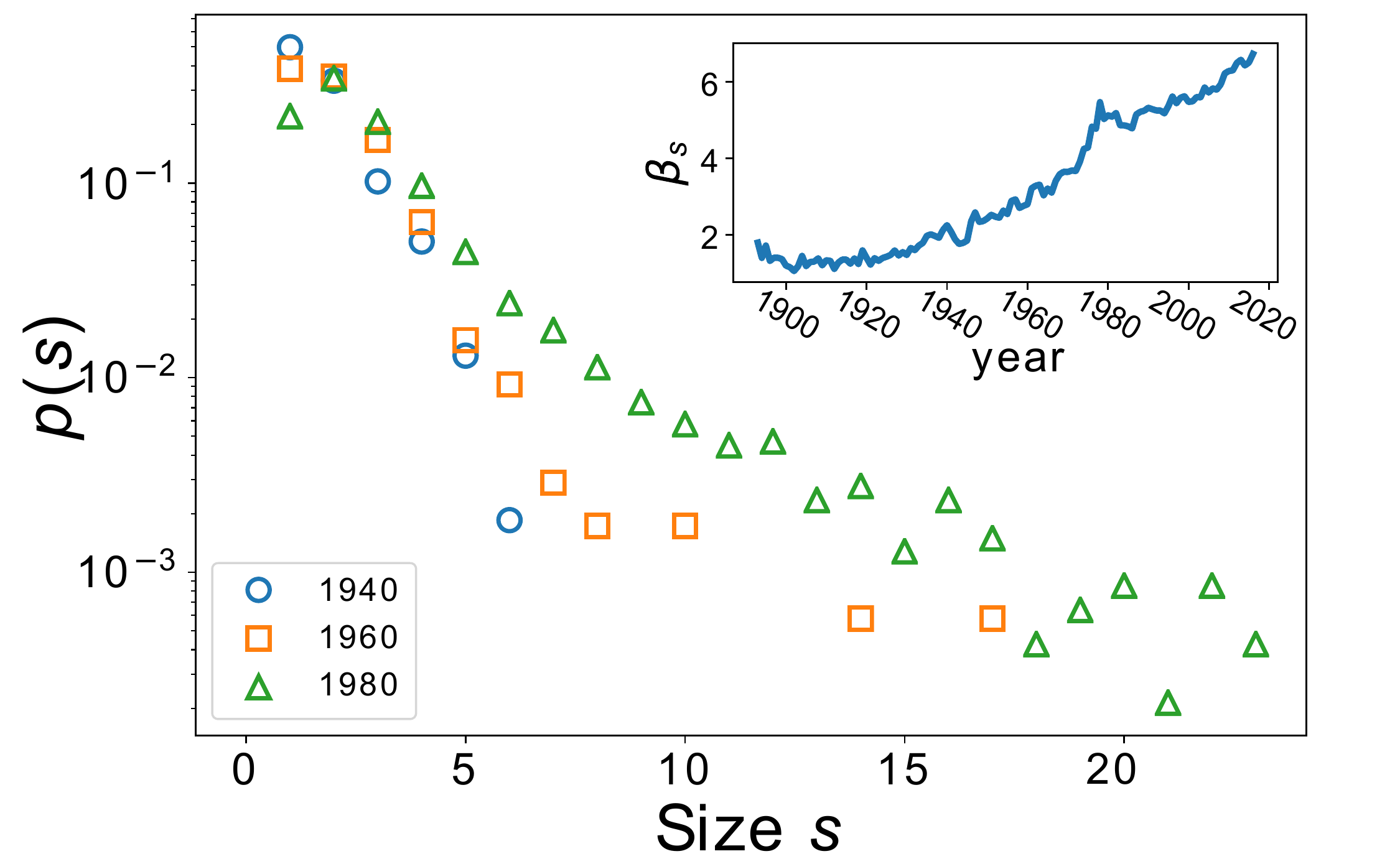}
\caption{\textbf{Distributions of the number $s$ of co-authors of a paper in the APS dataset}. The distribution becomes broader in more recent times. 
Inset: evolution of the ratio $\beta_s = \langle s^2 \rangle / \langle s \rangle$.
}
\label{fig:APS}
\end{figure}

\section{Simplices and simplicial complexes}
\label{sec:definitions}
We provide here several definitions related to simplicial complexes and their properties. A full treatment can be found in \cite{hatcher2005algebraic}. 

\begin{itemize}
\item In its simplest form, a $k$-dimensional simplex ($k$-simplex) $\sigma$ is a set of vertices $\sigma = [p_0, \ldots, p_{k-1}]$. The $d$-dimensional subfaces of $\sigma$ are all the subsets of $\sigma$ composed by $d+1$ vertices of $\sigma$.

\item A simplicial complex $X$ is a collection of simplices such that if $\sigma \in X$ then all its propers subfaces are also in $X$. 

\item The $k$-skeleton $X_k$ of $X$ is the simplicial complex obtained by the union of all the simplices in $X$ with dimension equal or lesser than $k$; the 1-skeleton of a simplicial complex is then the collection of all its edges, hence the underlying graph.

\item We define the set of $n$-dimensional chains $C_n(X)$ of a simplicial complex $X$ as the formal sums of $n$-simplices. This is formally expressed as:
\begin{equation}
C_n(X) =\{r_1\sigma_1 + r_2\sigma_2+ ... | r_i \in \mathbb{Z}, \sigma_i \in X_n\}.
\end{equation}

\item We define then the \emph{boundary map} $\partial_n$, which maps $n$-dimensional chains $C_n(X)$ to $(n-1)$-dimensional chains $C_{n-1}(X)$ and corresponds to the intuitive idea of mapping a simplex to its boundary. Formally the boundary map is defined as: 
\begin{eqnarray}
&\partial _{n}&:C_{n}(X) \to C_{n-1}(X) \\
& &\partial _{n}[v_0, ..., v_n]  = \sum _{i=0}^{n}(-1)^{i}[v_{0},\ldots ,{\hat {v}}_{i},\ldots ,v_{n}]
\end{eqnarray}
where we omit the vertex with the hat. For example, a $2$-simplex (a full triangle) is mapped to the alternated sum of its three concatenated edges ($1$-simplices).
The boundary map satisfies $\partial_{n}\partial_{n+1} = 0 \quad \forall n$
(hence $\text{ im }  \partial _{n+1} \subset \ker \partial _{n}$), which encodes the intuitive idea that the boundary of something has no boundary.  

\item The simplicial complex $X$ induces the \emph{chain complex}, $\cdots \rightarrow  C_{n+1} \rightarrow C_{n} \rightarrow C_{n-1} \rightarrow  \cdots$ through boundary maps $... \partial_{n+2}, \partial_{n+1}, \partial_{n}, \partial_{n-1}, ...$

\item The \emph{$n$-homology} group of $X$ is then defined by the quotient of two vector spaces, the kernel of the map $\partial_n$ 
quotiented by the image of the boundary map $\partial_{n+1}$,
\begin{align}
H_{n}(X)=\ker \partial _{n}/\text{ im } \partial _{n+1},
\end{align}
where $n$ indicates the dimension of the homology group. 
The corresponding Betti number $\beta_n$ is the dimension of $H_n$ and correspondingly the number of $n$-dimensional holes.

\item We define a simplicial connected $k$-component as a set of $k$-simplices in $X$ such that i) each simplex shares at least one $(k-1)$-face 
with another $k$-simplex in the component, and ii) given two simplices $A$ and $B$ in the component, there exists a connected path via $(k-1)$-adjacency 
(i.e., a list of  $k$-simplices $[A_0 = A,, A_1,\cdots,A_{n-1},A_n=B]$ such that for all $i$ $A_i$ and $A_{i+1}$ share a common  $(k-1)$-face)
between the two simplices (similarly to clique percolation \cite{palla2005uncovering}). 

\end{itemize}

We observe that, given a simplicial complex $X$, it is possible to consider two derived simplicial complexes:
\begin{itemize}
\item its $1$-skeleton, $X_1$, corresponding to the underlying graph (on which for example we study the spreading process of the SIS model);
\item the clique complex $Cl(X)$ built from $X_1$ by promoting every $(k+1)$-clique in $X_1$ to the corresponding $k$-simplex (e.g. a $4$-clique maps to a $3$-simplex). 
\end{itemize}
By construction, note that $X_1 \subset X \subset Cl(X)$: for example, while it is possible to have an empty triangle in $X$ 
composed by the simplices $[a,b], [a,c],[b,c]$, the same triangle would automatically be filled as $[a,b,c]$ in $Cl(X)$.

\section{Detailed computation of the aggregated degree}

The number of interactions of $i$ between times $0$ and
$T$ is given by:
\begin{eqnarray}
\kappa_T(i) = \bar{m}a_i T + \sum_{j\neq i} \frac{\bar{m}^2 T a_j}{N-1}
 \simeq \bar{m}T(a_i + \bar{m} \langle a \rangle)
 \label{eq_kappa}
 \end{eqnarray}
where the approximation holds for $N\gg 1$. 
We can then write $k_T(i)$ as \cite{Perra:2012}
\begin{eqnarray}
k_T(i) & = &  (N-1) \left[ 1 - (1 - \frac{1}{N-1})^{T\bar{m} (a_i + \ava \bar{m})} \right] \\
 &\simeq & (N-1) \left[ 1 -  e^{-\frac{T\bar{m} (a_i + \ava \bar{m})}{N-1}} \right] 
 \end{eqnarray}
 
Following \cite{Perra:2012}, it is also possible to derive the distribution of these degrees. 
We write $a_i = \eta x_i$, where $x_i$ is the activity potential and $\eta$ is a free parameter useful when data are available 
in order to match the average degree in the simulated temporal network to that of the real one. 
From this we can calculate the activity potential $x_i$ of node $i$ as:
\begin{eqnarray}
x_i = -\frac{N-1}{\bar{m}T\eta} \ln\left( 1 - \frac{k_T(i)}{N-1}\right)  - \frac{\bar{m}\langle a \rangle}{\eta} \\
\simeq \frac{k_T(i)}{\bar{m} T \eta}  - \frac{\bar{m}\langle a \rangle}{\eta}
\end{eqnarray}
where the approximation again holds for small $k/N$. 
Using the expressions above, by simple substitution we obtain the functional form for the integrated degree distribution:
\begin{eqnarray}
P_T(k) \sim \frac{1}{mT\eta} \frac{1}{1- \frac{k}{N-1}} F\left[-\frac{N-1}{mT\eta}\ln \left( 1- \frac{k}{N-1}\right) - \frac{m \langle a \rangle }{\eta}\right]  \ .
\end{eqnarray}

 For clarity of illustration, we also report here a larger version of Fig. 
 2a and 2b of the main text, 
 adding also in the inset the behaviour of $k_T$ as a function of the heterogeneity of the distribution of $s$,
 $\beta_s = \langle s^2 \rangle / \langle s \rangle$ (see inset of Fig. \ref{fig:aggregated-network-si}a): we 
 create SAD networks where we sample clique sizes from distributions $p(s)$ with fixed average and variable heterogeneities, in order
 to illustrate that,
if the simplex size s is allowed to fluctuate, the size of such fluctuations will affect the aggregated degree.

 \begin{figure}[tb]
  \centering
 \includegraphics[width=0.75\columnwidth]{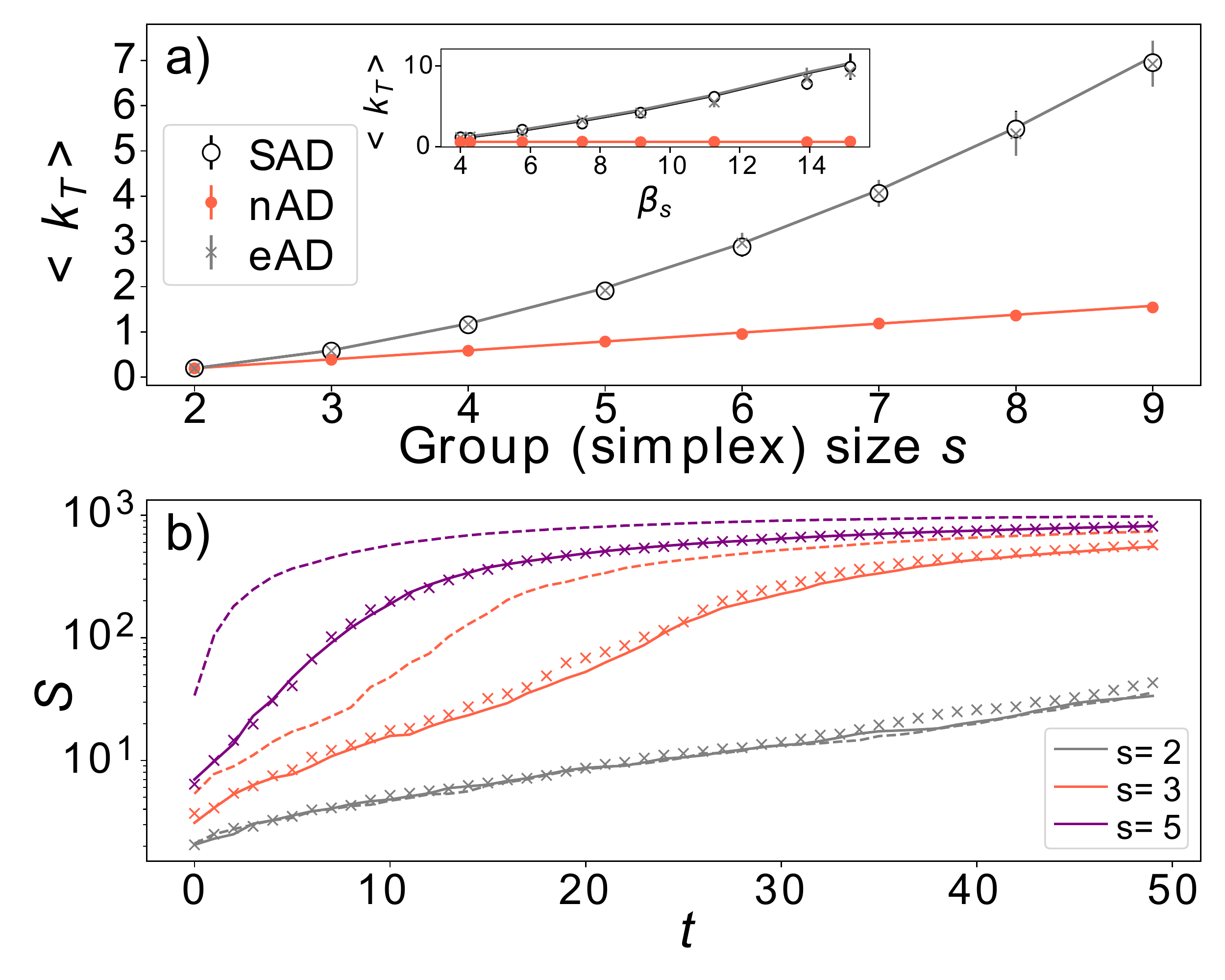}
 \caption{\textbf{Structural properties of SAD model.}
 (a) Average 
 aggregated degree $\langle k_T\rangle$ for the SAD model and corresponding nAD and eAD models for a range of simplex sizes $s$ (for $N=2000$, $T=10$, activities sampled from $P(a) = (a/a_0)^{-\alpha}$, $\alpha=2.1$ and $a_0=5\cdot 10^{-3}$). 
 $\langle k_T\rangle$ grows quadratically with $s$ in the SAD and in the eAD models and linearly in the nAD model. 
 Inset: $\langle k_T\rangle$ depends on the second moment of $s$
 in the SAD model (and in the eAD by inheritance of the fluctuations in $s$), but not in the nAD model.
 (b) Temporal growth of the size $S$ of the aggregated GCC for the SAD (solid lines), nAD (crosses) and eAD (dashed lines) for various fixed simplex size $s$. 
 The nAD and SAD models have the same behaviour as expected by design of the node-matched AD model, while the eAD model systematically generates much larger GCC sizes as compared to the corresponding SAD model.
 }
 \label{fig:aggregated-network-si}
 \end{figure}

\section{Number of $2$-simplices}

In a similar way as for the aggregated degree, one can compute 
$k_2(i,T)$ defined as the average number of $2$-simplices to which a node $i$ belongs, in the SAD aggregated until time $T$.

First, the average number of  $(s-1)$-simplices created between $0$ and $T$ to which $i$  belongs is computed, similarly to Eq. \eqref{eq_kappa}, as
\begin{equation}
\chi_T(i) = a_i T + \sum_{j \neq i} \frac{\bar{m} T a_j}{N-1} .
\end{equation}
Indeed, $i$ fires on average $a_i T$ times, and the second term is simply the average number of times that $i$ is chosen by another node $j$ to be in a $(s-1)$-simplex. 

Let us now fix $j$ and $k$, with $j\neq k$ and both distinct from $i$.
For each of the $\chi_T(i)$ $(s-1)$-simplices, the probability that $j$ and $k$ are part of it is $(s-1)(s-2)/((N-1)(N-2))$, so the probability that the pair $(j,k)$
is not part of any of these simplices is $p = (1 - (s-1)(s-2)/((N-1)(N-2)) )^{\chi_T(i)}$, and the probability that the pair is part of at least one common simplex
with $i$ is $1-p$. Each distinct pair $(j,k)$ part of at least one common simplex
with $i$ yields a distinct $2$-simplex $(i,j,k)$ to which $i$ belongs. 
As there are $(N-1)(N-2)/2$ pairs of nodes distinct from $i$, the number of $2$-simplices to which $i$ belongs is thus finally
\begin{equation}\label{eq:k2T}
k_2(i,T) = \binom{N-1}{2} \left( 
1 - \exp \left( - \frac{(s-1)(s-2)}{(N-1)(N-2)} T (a_i + (s-1) \langle a \rangle ) \right)  
\right)  .
\end{equation}

In addition to Figure 2d in the main text, Figures \eqref{fig:k2emp-vs-k2th} and \eqref{fig:k2-vs-T} confirm the validity of 
Eq. \eqref{eq:k2T} at a more detailed level. First, Fig. \eqref{fig:k2emp-vs-k2th} shows a scatterplot of the empirically measured value
of $k_2(i,T)$ vs. its theoretical value Eq. \eqref{eq:k2T} for each node $i$ in a SAD model with 
$N=2000$ nodes and activities sampled from a Pareto distribution with exponent $\alpha=2.4$. In Figure 
 \eqref{fig:k2emp-vs-k2th}, the empirical $k_2(i,T)$ is obtained by an average over $50$ realisations of the same SAD model
(i.e., each node $i$ has the same activity $a_i$ in the $50$ realisations). Note that the agreement between theory and empirical value
increases as $T$ increases and also for larger values of $k_2$, i.e., when the average is performed on larger numbers of events, showing that
the dispersion around the perfect agreement (given by the dashed diagonal line) is due to the finite number of realisations.

\begin{figure}
\centering
\includegraphics[width=.9\textwidth]{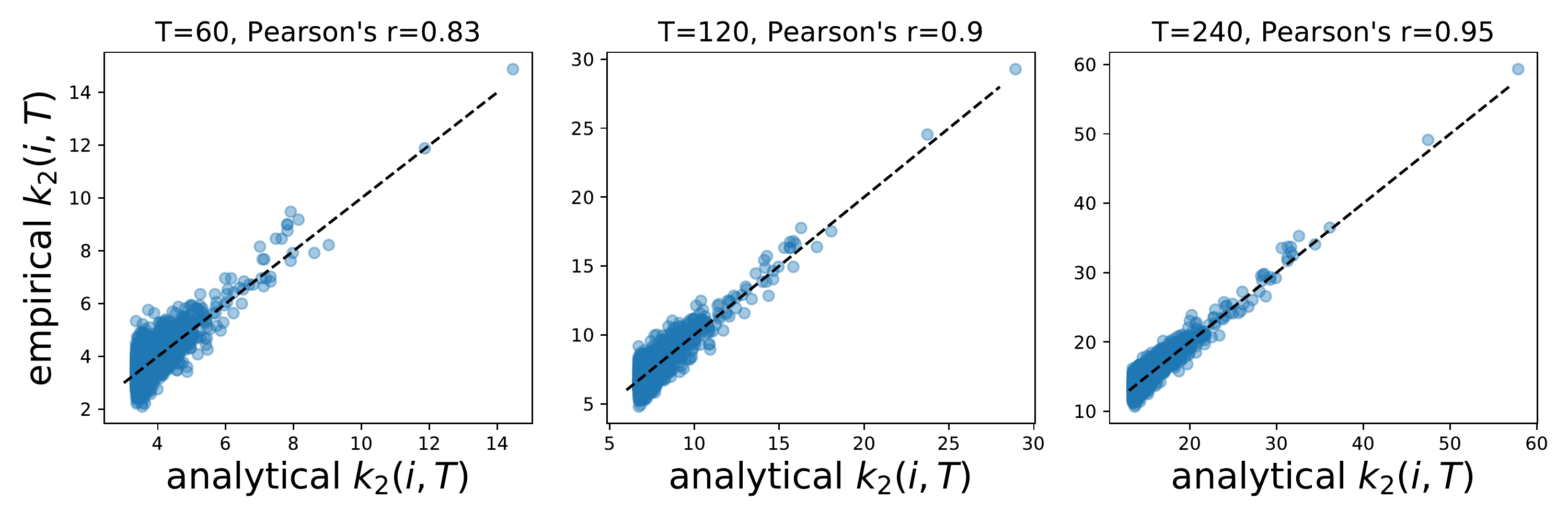}
\caption{\textbf{Theoretical versus empirical $k_2(i,T)$.} We show, as a scatterplot in which each dot corresponds to a node $i$,
the comparison of the predicted and the measured values of $k_2(i,T)$ for individual nodes $i$ 
(averaged over 50 iterations of the same SAD model) and for various aggregation times $T=60,120,240$ steps. 
The value of the  Pearson's correlation between theoretical and empirical values increases for longer aggregation times
(in all cases the correlation is significant, $p \ll  10^{-6}$). These results were obtained from a SAD model with $N=2000$ 
and activities sampled from a Pareto distribution with exponent $\alpha=2.4$.}\label{fig:k2emp-vs-k2th}
\end{figure}

\begin{figure}
\centering
\includegraphics[width=\textwidth]{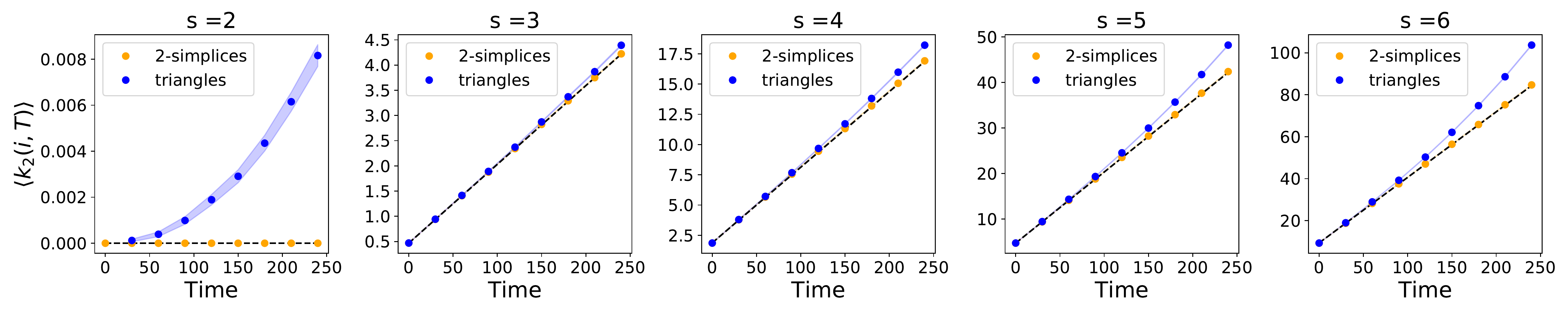}
\caption{\textbf{Temporal evolution of $\langle k_2(i,T) \rangle$.} We show the evolution of $\langle k_2(i,T)\rangle$ (i.e., 
$ k_2(i,T)$ averaged over nodes) as a function of the aggregation time $T$, for various values of $s$. 
Markers correspond to simulations, dashed lines to the prediction of Equation \eqref{eq:k2T}. 
As expected,  $\langle k_2(i,T)\rangle$ grows slower than the average number of all triangles 
(genuine $2$-simplices and  triangles composed only by three edges) to which a node belongs in the $1$-skeleton of the aggregated SAD. 
The latter corresponds to the only information available if one considers only the network structure, forfeiting the higher order simplicial one. 
Shaded areas represent Bayesian 95\% confidence intervals on $\langle k_2(i,T)\rangle$. The results were obtained from $200$ 
realizations of a SAD model with $N=2000$ and activities sampled from a Pareto distribution with exponent $\alpha=2.4$.}\label{fig:k2-vs-T}
\end{figure}

In addition, Fig. \eqref{fig:k2-vs-T} displays the average over nodes of $k_2(i,T)$ versus $T$, 
for various values of $s$, showing a perfect agreement between theory and numerical results. 
Importantly, Fig. \eqref{fig:k2-vs-T} also clearly confirms that $k_2(i,T)$
is different from (smaller than) the average number of triangles to which $i$ belongs in the aggregated network, as for instance a triangle $(i,j,k)$
in the aggregated network could be created by the aggregation
of three simplices created at different timesteps, containing respectively $(i,j)$ but not $k$, $(i,k)$ but not $j$ and $(j,k)$ but not $i$.

\begin{figure}
\centering
\includegraphics[width=.65\textwidth]{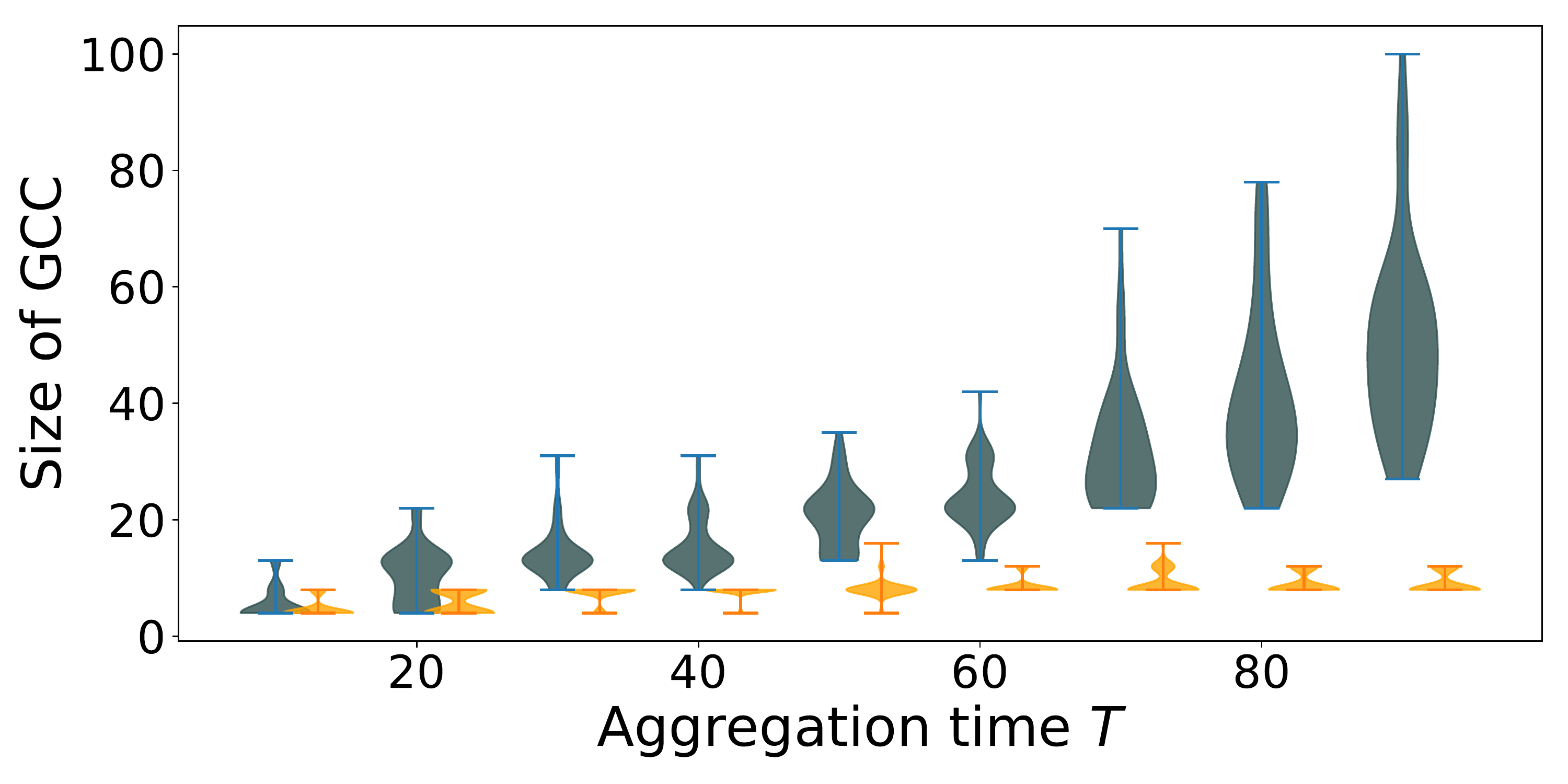}
\caption{\textbf{Temporal growth of largest simplicial and clique connected components.}
We show as violin plots, over a range of aggregation times, the distributions of the size of the largest connected $2$-simplex component (orange) in the 
aggregated SAD model and of the size of the largest $3$-clique component (dark gray) computed on the $1$-skeleton of the aggregated SAD model. 
As expected, the latter grows much faster than the former due to the presence of triangles that are not $2$-simplices. 
The results shown are obtained from $200$ iterations of the SAD model with $N=2000$ and activities sampled from a Pareto distribution with exponent $\alpha=2.4$.}\label{fig:si-simplex-gcc}
\end{figure}

Finally, Figure \eqref{fig:si-simplex-gcc} reports, as a function of the aggregation time $T$,  
the distributions of the size of the largest $2$-simplex component in the aggregated SAD model, 
and of the largest  $3$-clique component in the $1$-skeleton of the aggregated SAD model.
We recall (see Section \eqref{sec:definitions} for general definitions) that a connected $2$-simplex component is a set 
of $2$-simplices in the aggregated SAD such that i) each $2$-simplex shares at least one $1$-face (here, an edge) with another 
2-simplex in the component, and ii) given two simplices $A$ and $B$ in the component, there exists a connected path via $1$-simplex-adjacency 
between the two simplices, i.e., here, a list of $2$-simplices $[A_0=A,A_1,\cdots,A_{n-1},A_n=B]$ such that $\forall i$ $A_i$ and $A_{i+1}$ share a common edge.
For the  largest  $3$-clique component in the $1$-skeleton, the definition is the same but replacing $2$-simplices by triangles, 
as the $1$-skeleton does not distinguish between 
triangles and $2$-simplices. Hence, the number of possible paths is much larger in the $1$-skeleton and 
the size of the largest  $3$-clique component grows much faster than the one of the largest $2$-simplex component.

\section{Simplicial Laplacian}

The conventional graph Laplacian, $\mathcal{L}_0 = \mathbf{D} - \mathbf{A}$, is one of the  cornerstone of the study of how dynamical processes 
are affected by the underlying graph structure \cite{barrat2008dynamical}. $\mathcal{L}_0$ operates on the graph's nodes 
($0$-simplices): it is indeed an $N \times N$ matrix, where $N$ is the number of nodes.

In the case of a generic simplicial complex, it is possible to define a combinatorial (or simplicial) Laplacian $\mathcal{L}_k$ for each dimension $k$ of the simplices composing it. 
The definition of $\mathcal{L}_k$ \cite{horak2013spectra} is written in terms of boundary maps:
\begin{equation}
\mathcal{L}_k = \partial_{k+1} \partial^*_{k+1} + \partial^*_{k} \partial_{k}
\end{equation}
where it is easy to see that $\mathcal{L}_k$ maps $C_k \to C_k$. 
The two terms of the sum are also known respectively as $\mathcal{L}^{up}_k$ and $\mathcal{L}^{down}_k$. This is due to the fact that $\mathcal{L}^{up}_k$ goes up in dimension first and then down ($\mathcal{L}^{up}_k: C_k \to C_{k+1} \to C_k$), while the $\mathcal{L}^{down}_k$ does the opposite ($\mathcal{L}^{down}_k: C_k \to C_{k-1} \to C_k$). Finally, one can see that $\mathcal{L}_0$ reduces to $\mathcal{L}_0= \partial_{1} \partial^*_{1}$ since  $\partial^*_{0} \partial_{0}$ is always vanishing. \\
Similarly to the graph Laplacian, the properties of the Laplacian relate to the processes spreading on such simplices, for example as it pertains to random walks \cite{mukherjee2016random,parzanchevski2017simplicial} and control \cite{muhammad2006control}.  
Further, the dimension of $\ker \mathcal{L}_k$ can be proven to be the same as the dimension of the corresponding homology group $H_k$ \cite{friedman1998computing}. That is, the number of null eigenvalues is the same as the number of holes of the corresponding dimensions. This is true also for the standard graph Laplacian $\mathcal{L}_0$, where the number of null eigenvalues is the same as the number of connected components (0-dimensional holes). 

As $\mathcal{L}_0$ operates only on the nodes, its eigenspectrum is the same for the simplicial complex and its $1$-skeleton.
 However, this is no more true for larger values of $k$.
We can see an effect of this in the density of null eigenvalues of $\mathcal{L}_1$ for a simplicial complex $X$  when compared to its $1$-skeleton and to the clique complex built from its $1$-skeleton. 
Indeed, the 1-skeleton has no higher simplices, so all loops in the graph will count as holes, increasing the dimension of $H_1$ and hence of $\ker \mathcal{L}_1$. We therefore expect  $\beta_1^{skeleton}\geq \beta_1^{X}$. 
Conversely, the clique complex $Cl(X)$ will have by construction at least as many 2-simplices as $X$ itself and hence it will have $\beta_1^{Cl} \leq \beta_1^X$. 
In Figure 2d of the main text 
we confirm this numerically. 

\clearpage
\newpage

\section{SIS model on the SAD temporal network}
We consider the susceptible-infected-susceptible (SIS) model for disease spreading. In this model, nodes can be either susceptible (S) or infectious (I). Infectious individuals can propagate the disease to susceptible ones at rate $\beta$ whenever they are interacting, and recover spontaneously at rate $\mu$, becoming again susceptible.

We denote by $N_a$ the number of nodes with activity $a$.
The total number of nodes is $N = \int da N_a$.

$S_a^t$ and $I_a^t$ denote respectively the number of susceptible and infectious of activity $a$ at time $t$.
We have the simple conservation equation
$N_a = S_a^t + I_a^t$, and the total number of infectious
individuals is $I^t = \int da I_a^t$. 

\subsection{Case of a fixed clique size $s$.}
As explained in the main text, 
the time evolution of the number of infectious is given by the following equation:
\begin{eqnarray}
I_a^{t+\Delta t} - I_a^t  = - \mu \Delta t I_a^t  \nonumber
&+&\beta \Delta t S_a^t a (s-1) \int da' \frac{I_{a'}^t}{N} \\
&+& \beta \Delta t S_a^t \int da' a' \frac{I_{a'}^t}{N} (s-1)  \nonumber \\
&+& \beta \Delta t S_a^t \int da' a' \frac{S_{a'}^t}{N}(s-1) \times \int da'' \frac{I_{a''}^t}{N} (s-2)
\label{SIS_a_suppinfo}
\end{eqnarray}

If we integrate (\ref{SIS_a}) over $a$ we get an equation for $I_t$:
\begin{eqnarray}
I^{t+\Delta t} - I^t  &=& - \mu \Delta t I^t + \beta \Delta t (s-1) \ava I^t 
+ \beta \Delta t (s-1) \theta^t
+ \beta \Delta t (s-1)(s-2) \ava I^t \nonumber \\
&=& - \mu \Delta t I^t + \beta \Delta t (s-1)^2 \ava I^t 
+ \beta \Delta t (s-1) \theta^t
\end{eqnarray}
where $\theta^t = \int da a I_a^t$.

Multiplying (\ref{SIS_a}) by $a$ and integrating we get an equation for $\theta_t$:
\begin{eqnarray}
\theta^{t+\Delta t}  - \theta^t &=& -\mu \Delta t \theta^t 
+ \beta \Delta t (s-1) \asq I^t
+ \beta \Delta t (s-1) \ava \theta^t
+ \beta \Delta t (s-1)(s-2) \ava^2 I^t
\nonumber \\
&=&-\mu \Delta t \theta^t 
+ \beta \Delta t (s-1) \ava \theta^t
+ \beta \Delta t  \left( (s-1)\asq + (s-1)(s-2)\ava^2 \right) I^t
\end{eqnarray}

These equations can be rewritten as
\begin{equation} 
\left(
\begin{array}{c}
I^{t+\Delta t}  -I^t \\
\theta^{t+\Delta t} - \theta^t
\end{array}
\right)
= J
\left(
\begin{array}{c}
I^{t}   \\
\theta^{t}
\end{array}
\right)
\end{equation}
with
\begin{equation}
J = 
\left(
\begin{array}{cc}
-\mu + \beta (s-1)^2 \ava   &    \beta  (s-1)   \\
 \beta   \left( (s-1)\asq + (s-1)(s-2)\ava^2 \right)  & -\mu + \beta (s-1) \ava
\end{array}
\right)
\end{equation}

The characteristic polynomial of $J$ is
$$
(x+\mu)^2  -\beta s (s-1) \ava (x+\mu) + \beta^2 (s-1)^2 (\ava^2 - \asq)
$$
from which we find its eigenvalues:
$$
\Lambda_\pm = \left( \beta s (s-1) \ava -2\mu \pm \beta (s-1) \sqrt{s^2 \ava^2 + 4 (\asq - \ava^2)} \right)/2 .
$$

The epidemics does not vanish if and only if
 the largest eigenvalue $\Lambda_+$
is positive. This yields the epidemic threshold condition:
\begin{equation}
\frac{\beta}{\mu}  >  \lambda_c^{SAD}
\end{equation}
with 
\begin{equation}
 \lambda_c^{SAD} =\frac{2}{ s (s-1) \ava   + (s-1) \sqrt{s^2 \ava^2 + 4 (\asq - \ava^2)} }
\end{equation}


\subsection{Case of a distribution of clique sizes $P(s)$}
Let us assume that the size of the clique created
at each activation is taken from a distribution
$p(s)$. The size $s$ and the activity of the active
node at time $t$ are uncorrelated. 
With respect to the case of fixed clique size,
we then just add an integration on $p(s)$ in the evolution
equation of $I_a^t$:
\begin{eqnarray}
I_a^{t+\Delta t} - I_a^t  = - \mu \Delta t I_a^t  \nonumber
&+&\beta \Delta t \int ds p(s) S_a^t a (s-1) \int da' \frac{I_{a'}^t}{N} \\
&+& \beta \Delta t  \int ds p(s) S_a^t \int da' a' \frac{I_{a'}^t}{N} (s-1)  \nonumber \\
&+& \beta \Delta t \int ds p(s) S_a^t \int da' a' \frac{S_{a'}^t}{N}(s-1) \times \int da'' \frac{I_{a''}}{N} (s-2) .
\label{SIS_a_ps}
\end{eqnarray}

Proceeding as before, we obtain coupled equations
for $I^t$ and $\theta^t$, with the matrix J given now
by:
\begin{equation}
J = 
\left(
\begin{array}{cc}
-\mu + \beta \langle (s-1)^2 \rangle  \ava   &    \beta  \langle s-1 \rangle   \\
 \beta   \left( \langle s-1 \rangle \asq +  \langle (s-1)(s-2)\rangle\ava^2 \right)  & -\mu + \beta \langle s-1 \rangle \ava  .
\end{array}
\right)
\end{equation}

The eigenvalues of $J$ are now
$$
\Lambda_\pm = \left( \beta \langle s (s-1)\rangle \ava -2\mu 
\pm \beta  \sqrt{  \langle(s-1)(s-2) \rangle  \langle(s-1)(s+2) \rangle  \ava^2 + 4 \langle s-1 \rangle^2 \asq } \right)/2
$$
Imposing that the largest one
is positive gives the epidemic threshold condition
\begin{equation}
\frac{\beta}{\mu}  >  \lambda_c^{SAD} 
\end{equation}
with 
\begin{equation}
 \lambda_c^{SAD} =
 \frac{2}{ \langle s (s-1)\rangle \ava   +  
\sqrt{
\langle(s-1)(s-2) \rangle  \langle(s-1)(s+2) \rangle  \ava^2 + 4 \langle s-1 \rangle^2 \asq
}} .
\end{equation}

We observe that the epidemic threshold goes to $0$ as $1/\langle s^2 \rangle$ if the fluctuations of the clique size diverge, at fixed average
$\langle s \rangle$. We illustrate this in Fig. \eqref{fig:si-threshold-k2-dependence}.

\begin{figure}[h!]
\centering
\includegraphics[width=0.7\textwidth]{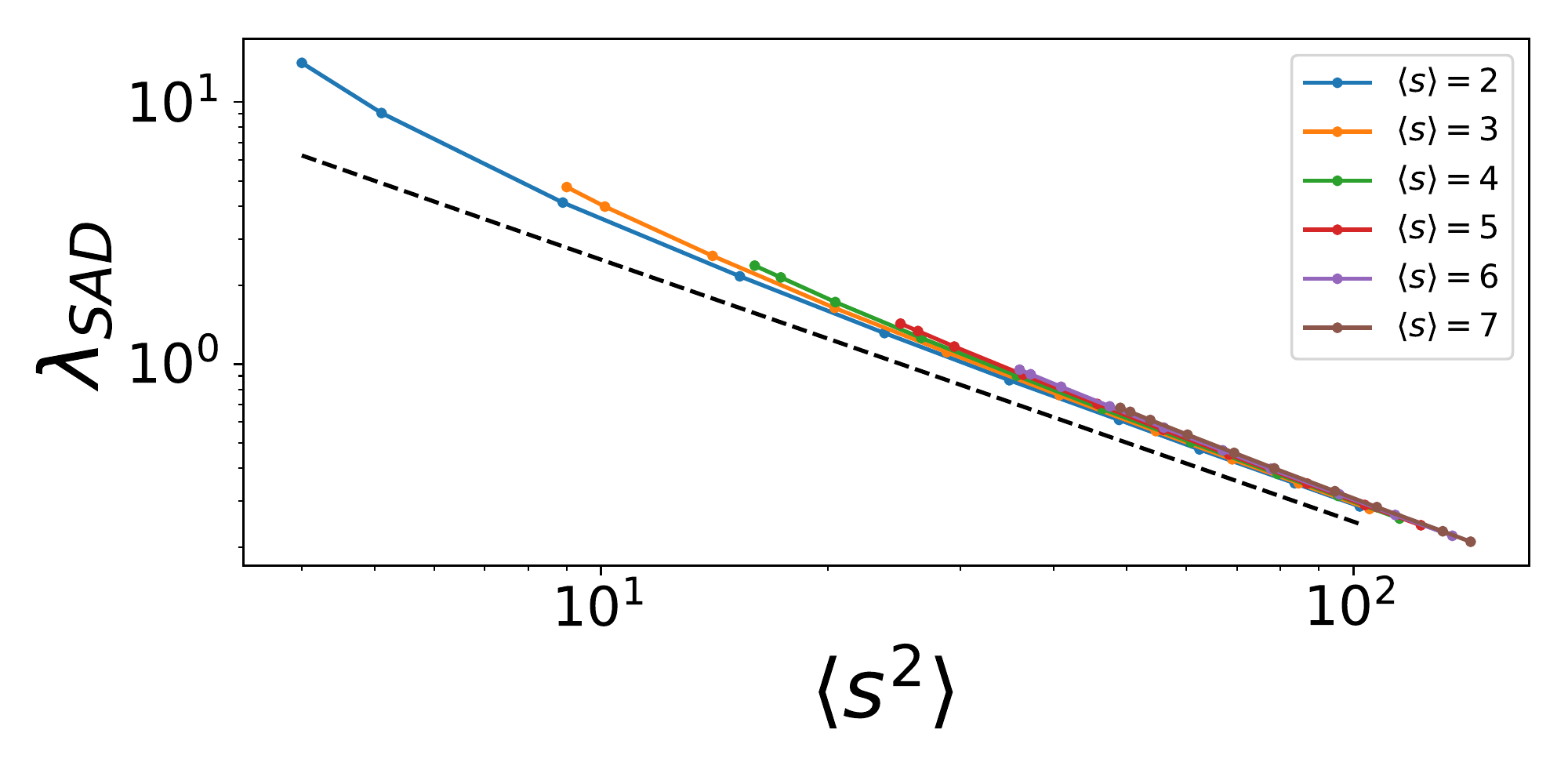}
\caption{\textbf{Dependence of $\lambda_{SAD}$ on $\langle s^2 \rangle$.} 
We show explicitly that $\lambda_{SAD}$ decreases proportionally to $\sim 1 / \langle s^2 \rangle$ for various average simplex sizes $\langle s \rangle$. We used $\langle a \rangle = 0.035$ and $\langle a^2 \rangle = 0.001$. The dashed black line is a reference to guide the eye and has slope -1. }
\label{fig:si-threshold-k2-dependence}
\end{figure}

\subsection{Ratio between SAD and eAD epidemic thresholds}
Using the expressions for $\lambda_c^{SAD}$ and $\lambda_c^{eAD}$, it is possible to write explicitly the expression for the ratio between the two quantities:
\begin{equation}
\frac{\lambda_c^{SAD}}{\lambda_c^{eAD}} = \frac{\langle a \rangle + \sqrt{\langle a^2 \rangle} }{\langle a \rangle + \sqrt{\langle a^2 \rangle} \sqrt{\Theta}} 
\label{eq:epithratio}
\end{equation}
with 
\begin{equation}
\Theta = \frac{4 \langle s-1 \rangle^2 }{\langle s(s-1)\rangle^2}  +  
 \frac{\langle (s-1)(s-2) \rangle \langle (s-1)(s+2) \rangle }{\langle s(s-1)\rangle^2} \cdot \frac{\langle a \rangle^2 }{\langle a^2\rangle}
\end{equation}
where for the eAD we used $m=s(s-1)/2$. While the expression above can take values smaller than 1, for meaningful values of the first two orders of $s$ and $a$ ($s \geq 2$, $\langle s^2 \rangle \geq \langle s \rangle^2$) it is always larger or equal to 1, implying that the SAD critical threshold is always higher than that of the corresponding eAD model.

As an example, Fig. \ref{fig:aps-ratios} shows the ratios obtained when using the activity and clique
size distributions measured in each year in the APS dataset: this ratio changes significantly between the data 
of 1900 and 2015, driven by changes in  the activity fluctuations $\beta_a$ until the 1950s and in
the co-authorship size fluctuations $\beta_s$ afterwards. 

\begin{figure}
\includegraphics[width=.5\columnwidth]{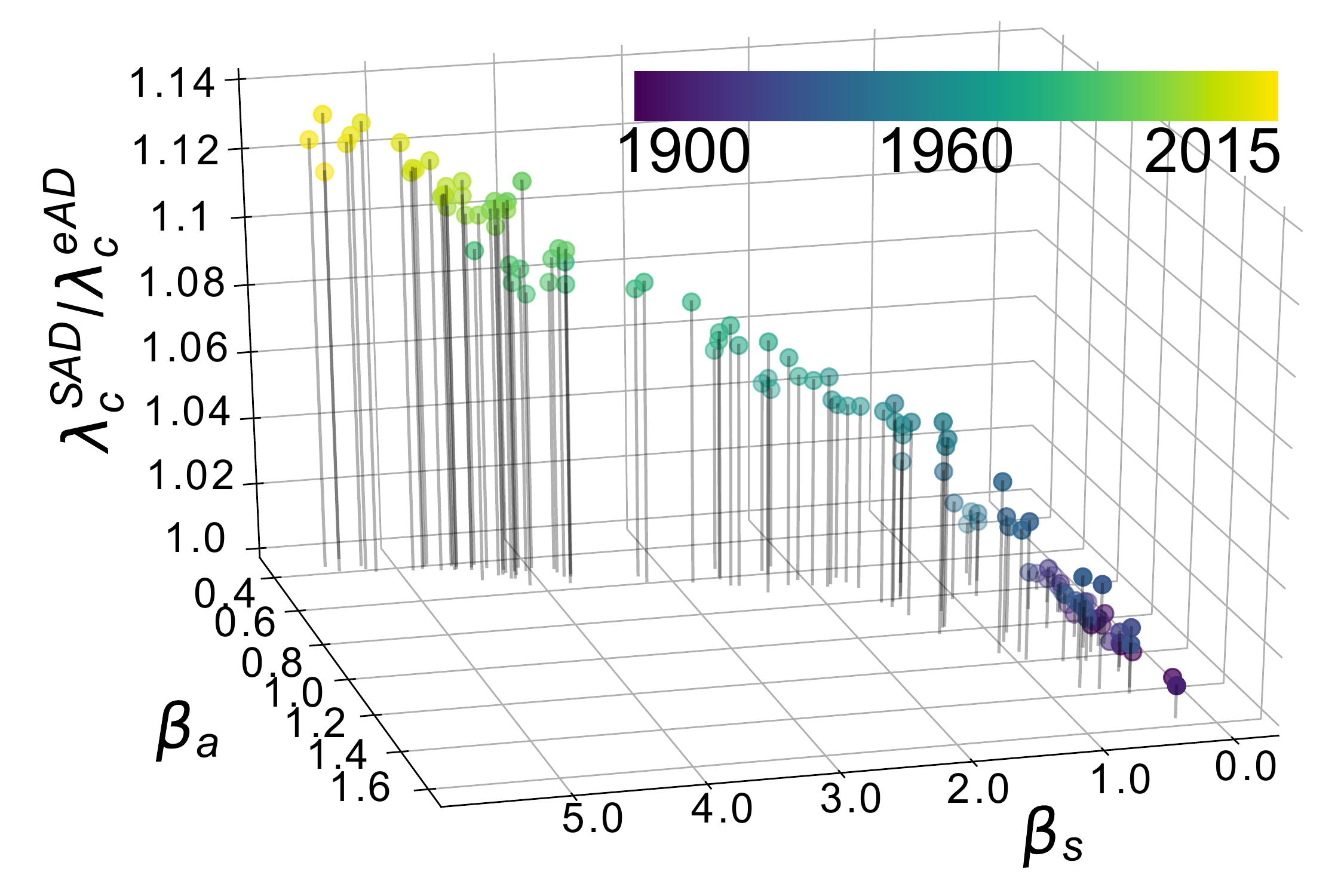}
\caption{\textbf{Ratio $\lambda_c^{SAD}/\lambda_c^{eAD}$ for co-authorship data.} 
We plot the ratio between the predicted epidemic thresholds
for the SAD and corresponding eAD models built using the 
empirical distributions $F(a)$ and $p(s)$ extracted each 
year from the  APS co-authorship data. 
The ratio grows between the years 1900 and 2015, 
driven by $\beta_a = \langle a^2 \rangle / \langle a \rangle$ until the 1950s and by $\beta_s=\langle s^2 \rangle / \langle s \rangle$  afterwards.}
\label{fig:aps-ratios}
\end{figure}

We also give in Table \ref{tab:SP-values} other examples of values obtained for the ratio between the predicted epidemic thresholds
for the SAD and corresponding eAD using data sets collected by the SocioPatterns
collaboration and publicly available \cite{SocioPatterns}: each data set describes the face-to-face contacts of individuals in a certain context (offices, conference,
hospital, school...) as measured by wearable sensors worn by the participants to the data collection, with a temporal resolution of $20$ seconds. We 
first aggregate the data on a certain temporal resolution $\Delta$, and measure on these data the distributions of node activities $F(a)$ and of clique
sizes $p(s)$. We then use Eq. \eqref{eq:epithratio} to compute the ratio between the predicted epidemic thresholds for SAD and eAD models defined
by these activity and clique size distributions. Note that the ratio increases for larger $\Delta$, due to the fact that the clique size distribution broadens
for smaller time resolution (larger $\Delta$).

 \begingroup
 \squeezetable
\begin{table}
\caption{}
\caption{Values of $\lambda_c^{SAD}/\lambda_c^{eAD}$ for SocioPatterns data: for each data set and for each temporal resolution, 
we extract the empirical distributions
$F(a)$ and $p(s)$, and compute the corresponding ratio of epidemic threshold using Eq. \eqref{eq:epithratio}.}
\label{tab:SP-values}
\begin{ruledtabular}
\begin{tabular}{lrrrrr}
{data set}  & context &    5min &   15min &   30min &   60min \\
\hline 
sThiers13   &  Highschool & 1.0220 &  1.0285 &  1.0339 &  1.0390 \\
sInVS13     &  Offices & 1.0085 &  1.0113 &  1.0127 &  1.0156 \\
sLH10       &  Hospital &1.0772 &  1.1118 &  1.1401 &  1.1748 \\
sSFHH       &  Conference &1.0583 &  1.0693 &  1.0831 &  1.1037 \\
sInVS15     & Offices & 1.0176 &  1.0262 &  1.0324 &  1.0468 \\
sLyonSchool & School & 1.0220 &  1.0353 &  1.0499 &  1.0700 \\
\end{tabular}
\end{ruledtabular}

\end{table}
\endgroup

\subsection{Case of a uniform activity}

Let us consider the case of a uniform activity: $a_i = a_0$ $\forall i$. Then $\langle a \rangle^2  = \langle a^2\rangle = a_0^2$ and it is easy to see
that 
\begin{equation}
  \lambda_c^{SAD} =\frac{1}{ s (s-1) a_0  } = \lambda_c^{eAD}
\end{equation}
i.e., the epidemic thresholds are then the same in the SAD and eAD models. To understand this point, it is useful to get back to the evolution equation
for the number of infectious individuals in each activity class, $I_a$. 
Since the activity is the same for all nodes, there is only one class of nodes, and we note $I = I_{a_0}$, $S=S_{a_0}$: the dynamical equation
will concern only this quantity, which averages over all nodes, and is thus mean-field. The intuition is that, as a result, the various terms concerning the various types of 
edges and contagion processes will be of a similar nature, no more depending on the activities of the joined nodes, and only the number of such 
possible events will be of importance: as SAD and eAD have the same number of created edges at each time step, the mean-field equations will be the same
for these two cases.

In mathematical details, all the integrals in the evolution
equation (\ref{SIS_a_suppinfo}) are simplified and one obtains
\begin{eqnarray}
I^{t+\Delta t} - I^t  = - \mu \Delta t I^t  \nonumber
&+&\beta \Delta t S^t a_0 (s-1)  \frac{I^t}{N} \\
&+& \beta \Delta t S^t a_0 \frac{I^t}{N} (s-1)  \nonumber \\
&+& \beta \Delta t S^t  a_0 \frac{S^t}{N}(s-1) \times  \frac{I}{N} (s-2)
\end{eqnarray}
yielding
\begin{equation}
I^{t+\Delta t} - I^t  = - \mu \Delta t I^t 
+\beta \Delta t a_0 (s-1)  \frac{S^t I^t}{N}  \left(   2 + \frac{(s-2)S^t}{N}  \right ) 
\end{equation}
At short times (or close to the threshold), $S^t \approx N$ so 
\begin{equation}
I^{t+\Delta t} - I^t  \approx - \mu \Delta t I^t 
+\beta \Delta t a_0 s(s-1)  \frac{S^t I^t}{N}  
\label{SIS_SAD_a0}
\end{equation}

In the eAD case, the evolution equation lacks the last term of the SAD model case:
\begin{eqnarray}
I_a^{t+\Delta t} - I_a^t  = - \mu \Delta t I_a^t  \nonumber
&+&\beta \Delta t S_a^t a m \int da' \frac{I_{a'}^t}{N} \\
&+& \beta \Delta t S_a^t \int da' a' \frac{I_{a'}^t}{N} m  \nonumber \\
\end{eqnarray}
and for a uniform activity $a_0$, the second and third terms are equal, yielding
\begin{equation}
I^{t+\Delta t} - I^t  = - \mu \Delta t I^t  
+2 \beta \Delta t S^t a_0 m \frac{I^t}{N} 
\end{equation}
which for $m=s(s-1)/2$ is the same as Eq. \eqref{SIS_SAD_a0}.

The equality between the evolution equations for the SIS on the SAD and eAD models when the activity is uniform can be understood as follows.
The various terms in the evolution equation correspond each to one type of links: for the eAD model, the term 
$\beta \Delta t S_a^t a m \int da' \frac{I_{a'}^t}{N}$ corresponds to links created by susceptible individuals of activity $a$ towards
infectious of any activity (hence the integral over $a'$), while the term
$\Delta t m S_a^t \int da' a' \frac{I_{a'}^t}{N}$ is instead due to infectious of activity $a$ creating links towards susceptibles of any activity. 
In other words, the $m$ links created at each step translate into $2m$ contagion opportunities, each weighted by the probabilities that the joined nodes
are one infectious and the other susceptible.

In the SAD model, the first terms are similar to the AD case, but the last term of the evolution equation,
$\beta \Delta t S_a^t \int da' a' \frac{S_{a'}^t}{N}(s-1) \times \int da'' \frac{I_{a''}^t}{N} (s-2)$, corresponds to links
joining nodes that are not active, of activity $a$ and $a''$, but brought together in contact by a node of activity $a'$ that creates a simplex.

When the activities are not uniform, each of these different terms is modulated in a different way by the activities of the linked nodes and weighted
by the probabilities of these nodes being respectively susceptible and infectious. 
As one focuses on nodes of activity $a$, one then integrates over the possible activities of the other nodes. 

When the activities are uniform instead, each link corresponds
potentially to a contagion in the same way as the other links, independently from the activities of the nodes it joins: the probability
that one is susceptible and the other infectious is the same for all links. Hence all these terms have in the end the same type
of contribution $S^t I^t/N$, and the factor is simply the number of links created ($m$ for the AD, $s(s-1)/2$ for the SAD). As the eAD has by definition the same
number of links as the SAD at each timestep, the number of contagion opportunities at each step are then the same and the evolution equations are the same. 
Obviously this holds for small $I^t$ for which one can ignore correlations due to the fact that the $s(s-1)/2$ of the SAD join only $s$ nodes.

\clearpage
\newpage

\section{Cascade model}
\subsection{Definition}
We consider here the model
of cascades introduced by D. Watts \cite{Watts:2002}
and generalized to temporal networks in Ref.~\cite{Karimi:2013}.
In this model, designed to represent the adoption of a behavior
or a product by individuals influenced by their social
contacts in a population, 
agents can be either in state $0$ (non-adopters) 
or $1$ (adopters). Agents in state $0$ can change state and
become adopters if a fraction of their social contacts
larger than $\phi \in [0,1]$ (the model parameter) are adopters.
In a static network framework, the social contacts are simply
the neighbors of an agent. Starting from a certain fraction of adopters, 
for instance taken at random among all agents, the agents' states 
evolve until no non-adopter has a fraction of adopters among
his/her neighbours larger than $\phi$: such a configuration is
blocked and the cascade stops. The efficiency of the cascade can then
be measured by the final fraction of adopters in the population.

Two important modifications need to be considered when the 
social contacts evolve in time, so that the adoption cascade
occurs along a temporal network of interactions:
\begin{itemize}
\item First, the model depends on another parameter, namely
the length $\theta$ of the time-window during which the social
contacts are considered: at time $t$, each 
non-adopter considers the time-window $(t-\theta,t]$ and becomes
adopter if the fraction of his/her number of contacts with adopters
among all interactions s/he had in this time-window 
is larger than $\phi$ ($\theta=1$ corresponds to taking into 
account only the interactions at time $t$). 

\item Second, as the contacts are changing over time,
a non-adopter could always have the possibility to make enough new
contacts 
with adopters to change state and become adopter: no configuration in which
some agents are still
non-adopters can be considered as blocked, and the cascade could
a priori continue until all agents are adopters. To measure the efficiency 
of the dynamics or compare various initial configurations or 
temporal networks, one should thus for instance fix a maximum time limit
for the cascade, or use 
empirical data of finite length \cite{Karimi:2013}.
\end{itemize}

While a full investigation of the model 
at varying $\phi$, $\theta$ on the one hand and at
different parameters
of the SAD temporal network on the other hand 
(either at fixed $s$ or with 
a distribution of sizes $p(s)$) is beyond our scope, 
we simply illustrate here some aspects of 
the complex phenomenology of the cascading processes.
In particular, we show that 
strong differences appear in the dynamical outcome of the 
process on SAD and AD temporal networks.

\subsection{Cascades on AD and SAD temporal networks}

Figure \ref{fig:cascade} shows the temporal evolution of the
average fraction of adopters in SAD temporal networks 
with fixed clique size $s$ 
and in corresponding eAD networks, for
$\theta=1$ and varying $\phi$. In the AD case, the dynamics becomes 
slower as $\phi$ increases, as can be expected since the fraction
of interactions with adopters needed to become an adopter increases, 
but the slowing down is very limited.
In the SAD model in contrast, as $\phi$ increases, the dynamics becomes extremely slow when $\phi$ increases. 

The strong difference between the 
dynamics on AD and SAD networks can 
be understood as follows. Let us first consider the AD case: at each time
step, if an adopter is activated and contacts a non-adopter, then the
fraction of adopters that the non-adopter sees is $1$, so s/he becomes
adopter, whatever the value of $\phi$. If instead the active node
is non-adopter, then s/he can become adopter if a fraction larger than 
$\phi$ of the $\binom{s}{2}$ agents s/he contacts are adopters. 
While the second mechanism becomes slower if $\phi$ increases, as at early
times it is less probable that a non-adopter nodes manages
to contact enough adopters to change state, 
the first mechanism
guarantees that the cascade can continue at any value of $\phi$.
The situation is quite different for the SAD model. Indeed, 
at each time step a clique of size $s$ is created. So each agent
has the same number $s-1$ of interactions, and a non-adopter
can become adopter only if the clique contains more 
than $(s-1)\phi$
adopters. For instance, if $s=5$, for $\phi=0.1$ a non-adopter becomes
adopter as soon as one adopter is present in the same clique. For
$\phi=0.3$ however the created clique needs to contain at least 
two adopters, for $\phi=0.5$ it needs $3$, etc: 
for $\phi \in [(n-1)/(s-1); n/(s-1)[$ a non-adopter needs 
to be put in interaction with at least $n$ adopters to change state. 
For any $n$ integer, the dynamics is thus much slower for
$\phi \ge n/(s-1)$ than for $\phi < n/(s-1)$.

\begin{figure*}
\includegraphics[width=.9\columnwidth]{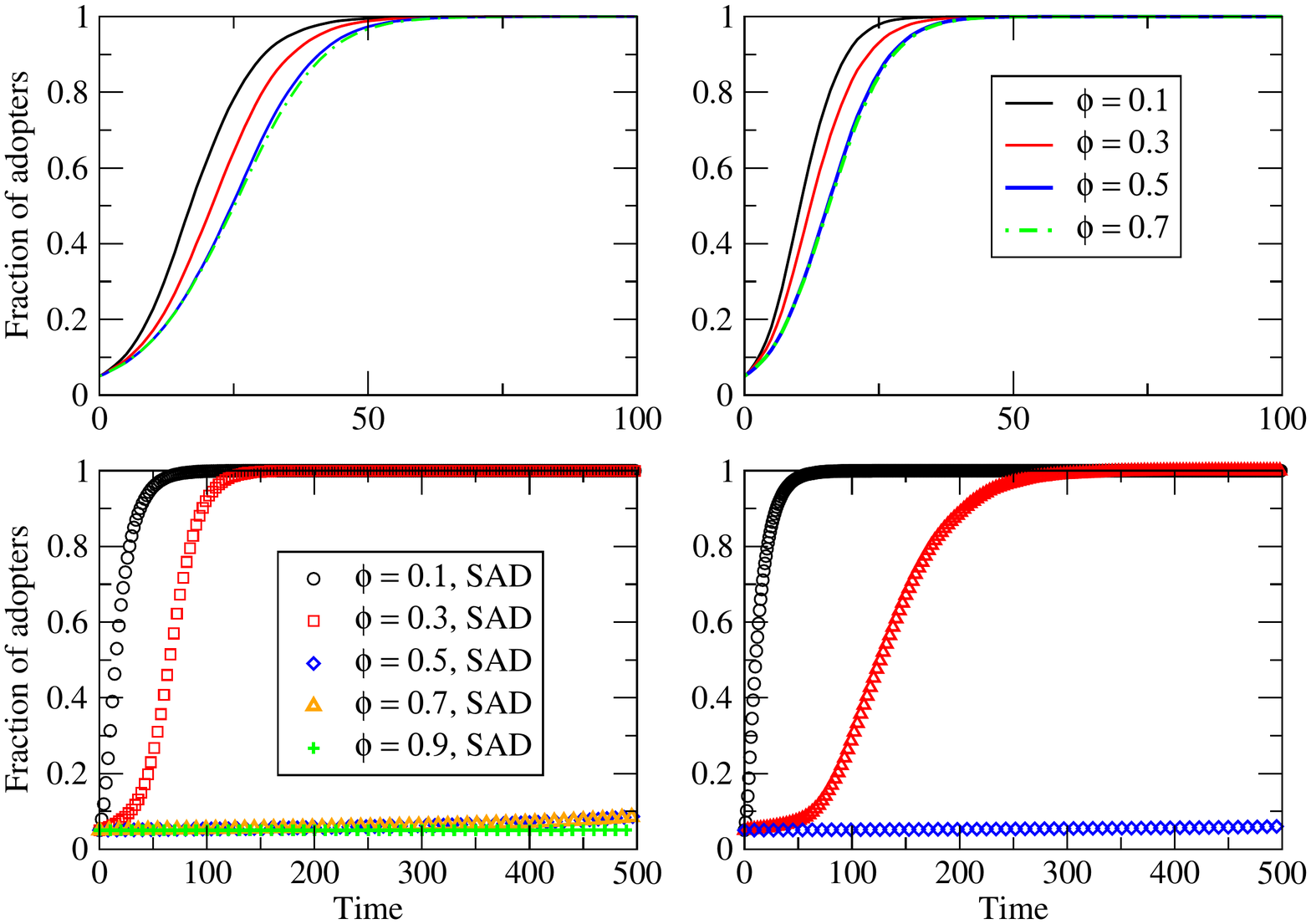}
\caption{Average fraction of adopters vs. time for cascades
unfolding along AD (top plots) and SAD (bottom plots) temporal networks
with $N = 500$ nodes, with
$s=5$ (left column) and $s=8$ (right column). Here
$\theta=1$.
}
\label{fig:cascade}
\end{figure*}

\subsection{Effect of the memory span $\theta$}

We have checked that this phenomenology is robust for $\theta > 1$.
As $\theta$ increases, the process becomes slower on 
both AD and SAD networks.  
This is in agreement with the results of \cite{Karimi:2013}, 
who find that the final cascade size  
on empirical temporal networks of finite duration
decreases when $\theta$
increases. This is due to the fact that the number of 
contacts increases when the time-window to 
consider $(t-\theta,t]$ becomes larger, but,
as the number of nodes in state 
$1$ is small at early times, their fraction during  $(t-\theta,t]$ 
typically decreases and thus it is more difficult for 
this fraction to be above $\phi$.

\subsection{Effect of the parameter $s$}

Figure \ref{fig:cascade_bis} compares the results of 
simulations of cascades on AD and SAD temporal networks
for various values of $s$ at fixed values of $\phi$,
with $\theta = 1$.

For the AD model, as $s$ increases, the cascade dynamics 
gets faster. This is the result of
a competition between 2 phenomena: 
at fixed $\phi$, if $s$ increases, the probability for an 
active node in status $0$ to have a fraction at least $\phi$ 
of its neighbors in state $1$ is smaller, at least at short times.
This would slow down the cascade.
However, an active node in status $1$ is able to
spread its opinion to all the nodes it contacts:
in the eAD model, a single active node can change the state of
$\binom{s}{2}$ other nodes, a number growing quadratically 
with $s$, so that this effect dominates and the cascade
becomes overall faster.

For the SAD model, the picture is more complicated.
If $s$ increases but the value of $\phi$
remains in $[(n-1)/(s - 1), n/(s - 1)[$ with fixed $n$, then 
with increased $s$ the spread becomes faster. Equivalently,
this corresponds to $s$ increasing within 
$[1 + (n-1)/\phi, 1 + n/\phi[$. When $s$ crosses
the value $1 + n/\phi$ however, the number of 
nodes in state $1$ needed to convince the other nodes
in the cliques increases by $1$, so the dynamics becomes
much slower. 

Let us take concrete examples. For $\phi = 0.1$, increasing
$s$ leads to faster dynamics until $s=11$. For 
$\phi = 0.3$, the intervals $[1 + n/\phi, 1 + (n + 1)/\phi[$
to consider are $[1,4.33[$, $[4.33,7.67[$,
$[7.67,11[$, etc. So $s=4$ leads to faster
dynamics than $s=3$ but $s=5$ becomes slower
at short times, $s=6$ is faster than $s=5$ and $s=8$ becomes
much slower ($n=3$ is needed).

For $\phi = 0.5$, the intervals $[1 + (n - 1)/\phi, 1 + n/\phi[$
to consider are $[1,3[$, $[3,5[$, $[5,7[$, $[7,9[$, etc.
So $s=4$ leads to faster
dynamics than $s=3$, $s=5$ is instead much slower,
etc.

Overall, the cascade velocity is non-monotonic with
respect to an increase in $s$ for the dynamics on the SAD
temporal network.

\begin{figure*}
\includegraphics[width=.9\columnwidth]{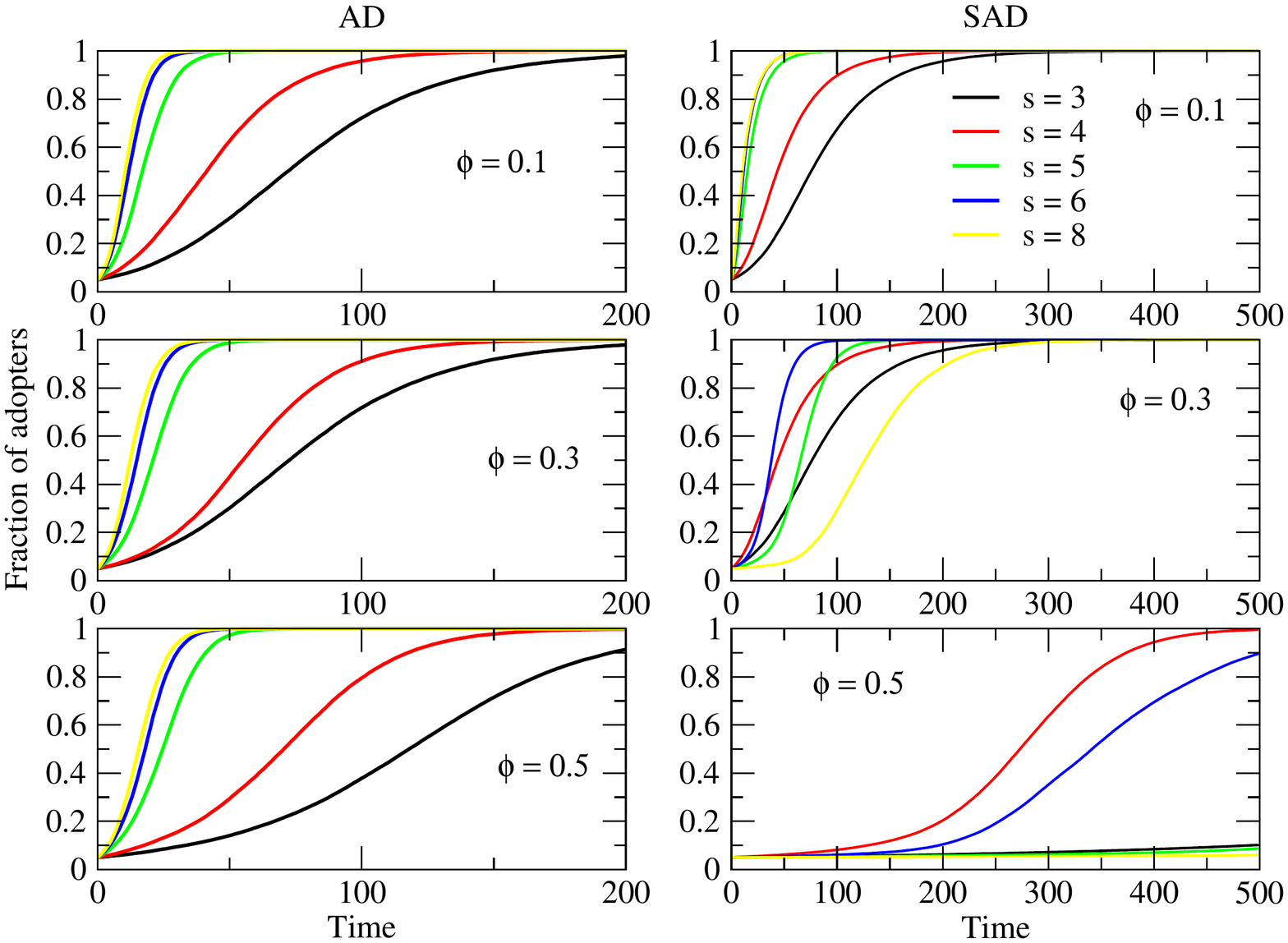}
\caption{Average fraction of adopters vs. time for cascades
unfolding along eAD (left plots) and SAD (right plots) temporal networks with $N = 500$ nodes, for various
values of $s$ and $\phi$. Here $\theta=1$.
}
\label{fig:cascade_bis}
\end{figure*}

\subsection{Effect of the variability in clique size}

We finally mention the effect of variability in the
values of $s$. Let us assume a distribution $p(s)$ centered
on a value $s_0$ and with tunable fluctuations around $s_0$,
quantified by a coefficient of variation $v$
(ratio between standard deviation and average).
On the AD model, an increased variability
simply speeds up slightly the cascading process
(see Fig.~\ref{fig:cascade_AD_v}).
On the other hand, a much more dramatic effect with very strong
speed-up of the dynamics can be obtained when the dynamics
takes place on the SAD model, as illustrated
in Fig.~\ref{fig:cascade_SAD_v}. Indeed, let us assume that 
$\phi \in [(n-1)/(s_0-1); n/(s_0-1)[$ with $n$ not too small, 
meaning that the dynamics with fixed $s_0$ is slow 
because a node in state $0$ needs to be in contact with at
least $n$ nodes in state $1$ to change state.
Variability in $s$ means that the clique formed can be 
sometimes smaller than $s_0$, and $\phi$ could
sometimes be smaller than $(n-1)/(s-1)$, in which case
only $n-1$ nodes in state $1$ are required to change state, and
the process becomes thus much faster.

\begin{figure*}
\includegraphics[width=.75\columnwidth]{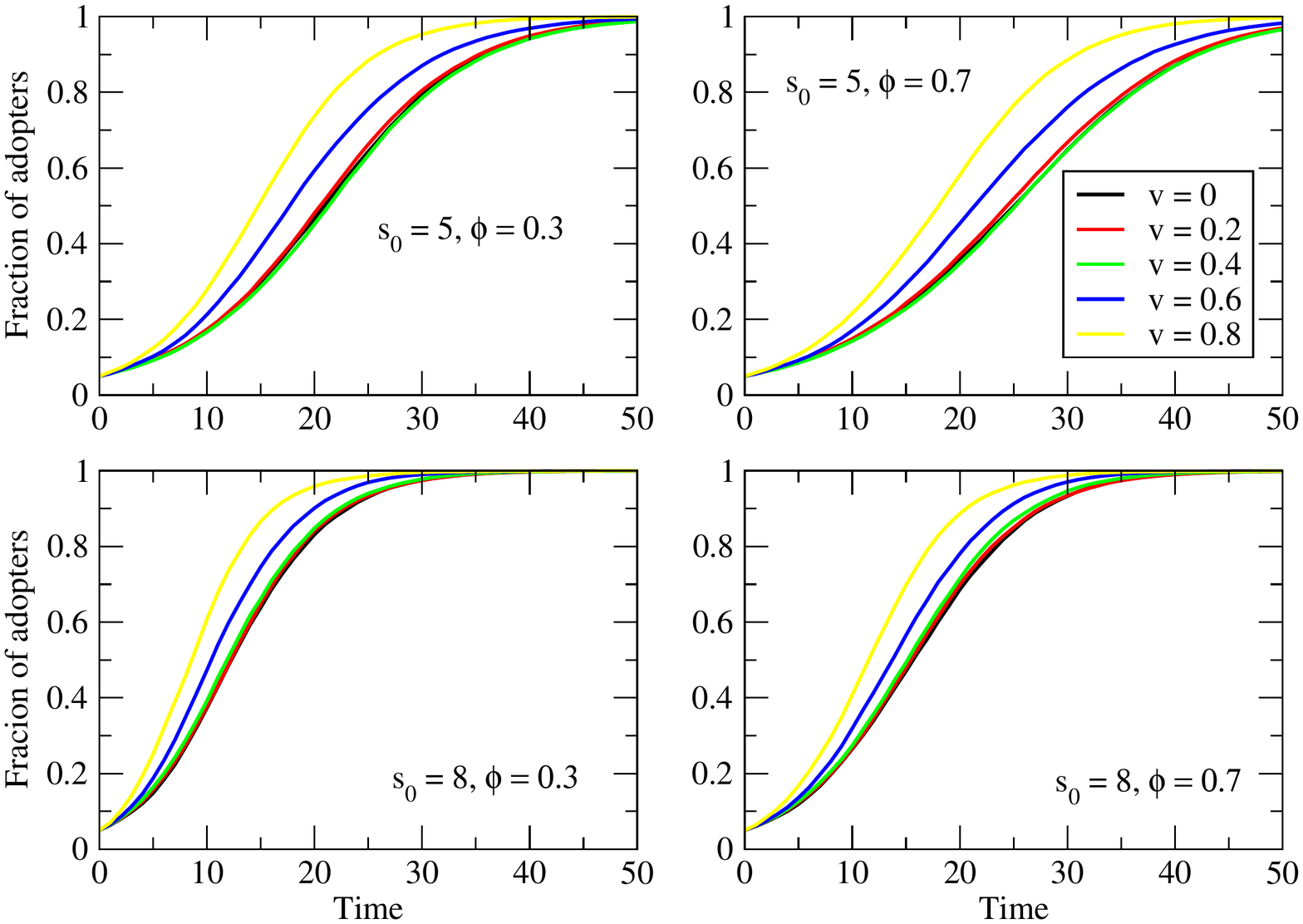}
\caption{Average fraction of adopters vs. time 
for cascades unfolding along eAD  
temporal networks with $N = 500$ nodes, for various
values of $s_0$ (average of $p(s)$), $v$ (coefficient of variation of $p(s)$) and $\phi$. Here $\theta=1$.
}
\label{fig:cascade_AD_v}
\end{figure*}

\begin{figure*}
\includegraphics[width=.75\columnwidth]{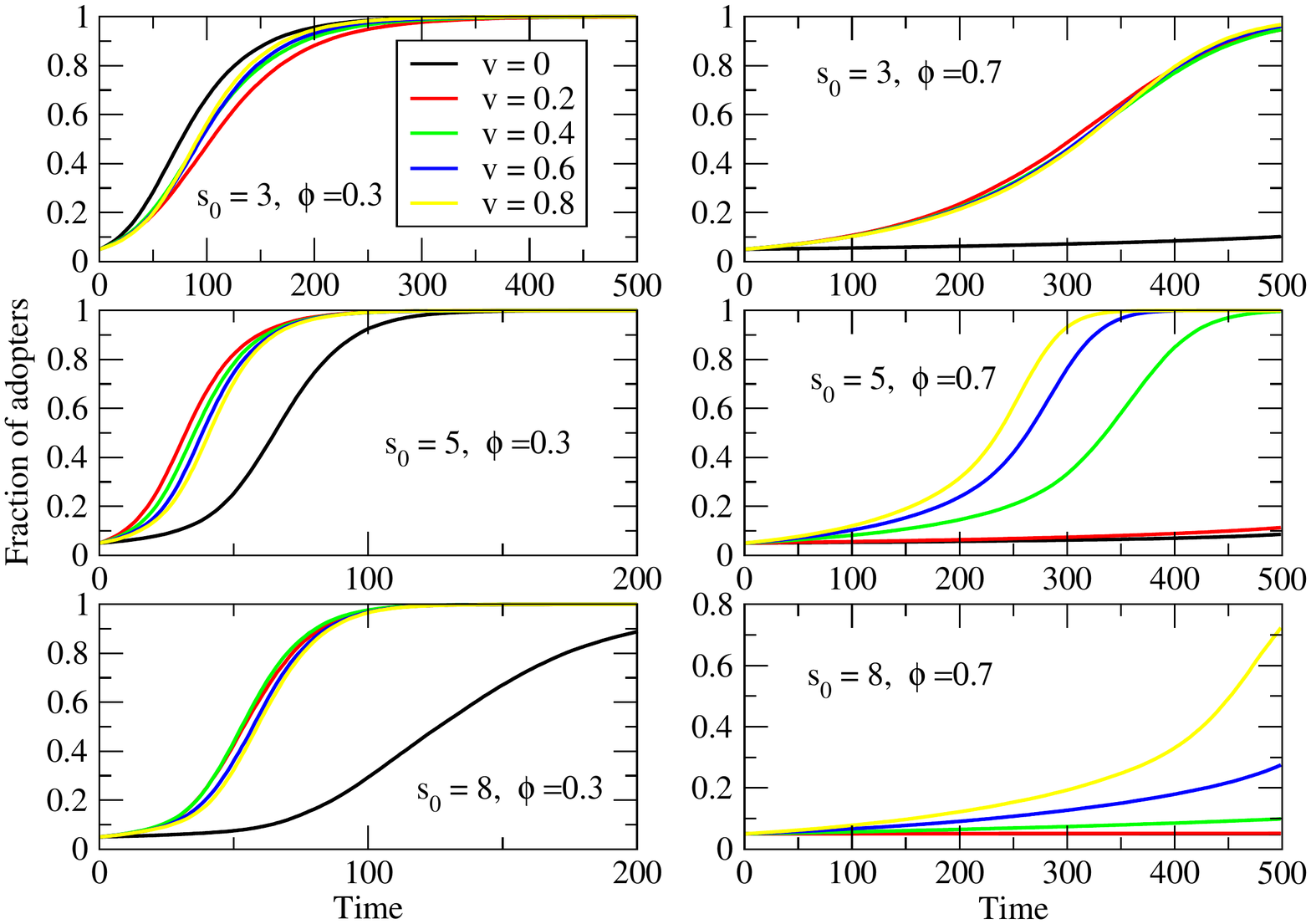}
\caption{Average fraction of adopters vs. time for cascades
unfolding along SAD temporal networks with 
$N = 500$ nodes, for various
values of $s_0$ (average of $p(s)$), $v$ (coefficient of variation of $p(s)$) and $\phi$. Here $\theta=1$.
}
\label{fig:cascade_SAD_v}
\end{figure*}

\end{widetext}


\begin{thebibliography}{34}%
\makeatletter
\providecommand \@ifxundefined [1]{%
 \@ifx{#1\undefined}
}%
\providecommand \@ifnum [1]{%
 \ifnum #1\expandafter \@firstoftwo
 \else \expandafter \@secondoftwo
 \fi
}%
\providecommand \@ifx [1]{%
 \ifx #1\expandafter \@firstoftwo
 \else \expandafter \@secondoftwo
 \fi
}%
\providecommand \natexlab [1]{#1}%
\providecommand \enquote  [1]{``#1''}%
\providecommand \bibnamefont  [1]{#1}%
\providecommand \bibfnamefont [1]{#1}%
\providecommand \citenamefont [1]{#1}%
\providecommand \href@noop [0]{\@secondoftwo}%
\providecommand \href [0]{\begingroup \@sanitize@url \@href}%
\providecommand \@href[1]{\@@startlink{#1}\@@href}%
\providecommand \@@href[1]{\endgroup#1\@@endlink}%
\providecommand \@sanitize@url [0]{\catcode `\\12\catcode `\$12\catcode
  `\&12\catcode `\#12\catcode `\^12\catcode `\_12\catcode `\%12\relax}%
\providecommand \@@startlink[1]{}%
\providecommand \@@endlink[0]{}%
\providecommand \url  [0]{\begingroup\@sanitize@url \@url }%
\providecommand \@url [1]{\endgroup\@href {#1}{\urlprefix }}%
\providecommand \urlprefix  [0]{URL }%
\providecommand \Eprint [0]{\href }%
\providecommand \doibase [0]{http://dx.doi.org/}%
\providecommand \selectlanguage [0]{\@gobble}%
\providecommand \bibinfo  [0]{\@secondoftwo}%
\providecommand \bibfield  [0]{\@secondoftwo}%
\providecommand \translation [1]{[#1]}%
\providecommand \BibitemOpen [0]{}%
\providecommand \bibitemStop [0]{}%
\providecommand \bibitemNoStop [0]{.\EOS\space}%
\providecommand \EOS [0]{\spacefactor3000\relax}%
\providecommand \BibitemShut  [1]{\csname bibitem#1\endcsname}%
\let\auto@bib@innerbib\@empty
\bibitem [{\citenamefont {Albert}\ and\ \citenamefont
  {Barab\'asi}(2002)}]{Albert:2002}%
  \BibitemOpen
  \bibfield  {author} {\bibinfo {author} {\bibfnamefont {R.}~\bibnamefont
  {Albert}}\ and\ \bibinfo {author} {\bibfnamefont {A.-L.}\ \bibnamefont
  {Barab\'asi}},\ }\href@noop {} {\bibfield  {journal} {\bibinfo  {journal}
  {Rev. Mod. Phys.}\ }\textbf {\bibinfo {volume} {74}},\ \bibinfo {pages} {47 }
  (\bibinfo {year} {2002})}\BibitemShut {NoStop}%
\bibitem [{\citenamefont {Barrat}\ \emph
  {et~al.}(2008{\natexlab{a}})\citenamefont {Barrat}, \citenamefont
  {Barthelemy},\ and\ \citenamefont {Vespignani}}]{BBV}%
  \BibitemOpen
  \bibfield  {author} {\bibinfo {author} {\bibfnamefont {A.}~\bibnamefont
  {Barrat}}, \bibinfo {author} {\bibfnamefont {M.}~\bibnamefont {Barthelemy}},
  \ and\ \bibinfo {author} {\bibfnamefont {A.}~\bibnamefont {Vespignani}},\
  }\href@noop {} {\emph {\bibinfo {title} {Dynamical processes on complex
  networks}}}\ (\bibinfo  {publisher} {Cambridge University Press
  (Cambridge)},\ \bibinfo {year} {2008})\BibitemShut {NoStop}%
\bibitem [{\citenamefont {Ramasco}\ \emph {et~al.}(2004)\citenamefont
  {Ramasco}, \citenamefont {Dorogovtsev},\ and\ \citenamefont
  {Pastor-Satorras}}]{Ramasco:2004}%
  \BibitemOpen
  \bibfield  {author} {\bibinfo {author} {\bibfnamefont {J.~J.}\ \bibnamefont
  {Ramasco}}, \bibinfo {author} {\bibfnamefont {S.~N.}\ \bibnamefont
  {Dorogovtsev}}, \ and\ \bibinfo {author} {\bibfnamefont {R.}~\bibnamefont
  {Pastor-Satorras}},\ }\href {\doibase 10.1103/PhysRevE.70.036106} {\bibfield
  {journal} {\bibinfo  {journal} {Phys. Rev. E}\ }\textbf {\bibinfo {volume}
  {70}},\ \bibinfo {pages} {036106} (\bibinfo {year} {2004})}\BibitemShut
  {NoStop}%
\bibitem [{\citenamefont {Giusti}\ \emph {et~al.}(2015)\citenamefont {Giusti},
  \citenamefont {Pastalkova}, \citenamefont {Curto},\ and\ \citenamefont
  {Itskov}}]{giusti2015clique}%
  \BibitemOpen
  \bibfield  {author} {\bibinfo {author} {\bibfnamefont {C.}~\bibnamefont
  {Giusti}}, \bibinfo {author} {\bibfnamefont {E.}~\bibnamefont {Pastalkova}},
  \bibinfo {author} {\bibfnamefont {C.}~\bibnamefont {Curto}}, \ and\ \bibinfo
  {author} {\bibfnamefont {V.}~\bibnamefont {Itskov}},\ }\href@noop {}
  {\bibfield  {journal} {\bibinfo  {journal} {Proceedings of the National
  Academy of Sciences}\ }\textbf {\bibinfo {volume} {112}},\ \bibinfo {pages}
  {13455} (\bibinfo {year} {2015})}\BibitemShut {NoStop}%
\bibitem [{\citenamefont {Reimann}\ \emph {et~al.}(2017)\citenamefont
  {Reimann}, \citenamefont {Nolte}, \citenamefont {Scolamiero}, \citenamefont
  {Turner}, \citenamefont {Perin}, \citenamefont {Chindemi}, \citenamefont
  {D{\l}otko}, \citenamefont {Levi}, \citenamefont {Hess},\ and\ \citenamefont
  {Markram}}]{reimann2017cliques}%
  \BibitemOpen
  \bibfield  {author} {\bibinfo {author} {\bibfnamefont {M.~W.}\ \bibnamefont
  {Reimann}}, \bibinfo {author} {\bibfnamefont {M.}~\bibnamefont {Nolte}},
  \bibinfo {author} {\bibfnamefont {M.}~\bibnamefont {Scolamiero}}, \bibinfo
  {author} {\bibfnamefont {K.}~\bibnamefont {Turner}}, \bibinfo {author}
  {\bibfnamefont {R.}~\bibnamefont {Perin}}, \bibinfo {author} {\bibfnamefont
  {G.}~\bibnamefont {Chindemi}}, \bibinfo {author} {\bibfnamefont
  {P.}~\bibnamefont {D{\l}otko}}, \bibinfo {author} {\bibfnamefont
  {R.}~\bibnamefont {Levi}}, \bibinfo {author} {\bibfnamefont {K.}~\bibnamefont
  {Hess}}, \ and\ \bibinfo {author} {\bibfnamefont {H.}~\bibnamefont
  {Markram}},\ }\href@noop {} {\bibfield  {journal} {\bibinfo  {journal}
  {Frontiers in computational neuroscience}\ }\textbf {\bibinfo {volume}
  {11}},\ \bibinfo {pages} {48} (\bibinfo {year} {2017})}\BibitemShut {NoStop}%
\bibitem [{\citenamefont {Patania}\ \emph {et~al.}(2017)\citenamefont
  {Patania}, \citenamefont {Petri},\ and\ \citenamefont
  {Vaccarino}}]{patania2017shape}%
  \BibitemOpen
  \bibfield  {author} {\bibinfo {author} {\bibfnamefont {A.}~\bibnamefont
  {Patania}}, \bibinfo {author} {\bibfnamefont {G.}~\bibnamefont {Petri}}, \
  and\ \bibinfo {author} {\bibfnamefont {F.}~\bibnamefont {Vaccarino}},\
  }\href@noop {} {\bibfield  {journal} {\bibinfo  {journal} {EPJ Data Science}\
  }\textbf {\bibinfo {volume} {6}},\ \bibinfo {pages} {18} (\bibinfo {year}
  {2017})}\BibitemShut {NoStop}%
\bibitem [{\citenamefont {Bianconi}\ and\ \citenamefont
  {Rahmede}(2016)}]{bianconi2016network}%
  \BibitemOpen
  \bibfield  {author} {\bibinfo {author} {\bibfnamefont {G.}~\bibnamefont
  {Bianconi}}\ and\ \bibinfo {author} {\bibfnamefont {C.}~\bibnamefont
  {Rahmede}},\ }\href@noop {} {\bibfield  {journal} {\bibinfo  {journal}
  {Physical Review E}\ }\textbf {\bibinfo {volume} {93}},\ \bibinfo {pages}
  {032315} (\bibinfo {year} {2016})}\BibitemShut {NoStop}%
\bibitem [{\citenamefont {Bianconi}\ \emph {et~al.}(2015)\citenamefont
  {Bianconi}, \citenamefont {Rahmede},\ and\ \citenamefont
  {Wu}}]{bianconi2015complex}%
  \BibitemOpen
  \bibfield  {author} {\bibinfo {author} {\bibfnamefont {G.}~\bibnamefont
  {Bianconi}}, \bibinfo {author} {\bibfnamefont {C.}~\bibnamefont {Rahmede}}, \
  and\ \bibinfo {author} {\bibfnamefont {Z.}~\bibnamefont {Wu}},\ }\href@noop
  {} {\bibfield  {journal} {\bibinfo  {journal} {Physical Review E}\ }\textbf
  {\bibinfo {volume} {92}},\ \bibinfo {pages} {022815} (\bibinfo {year}
  {2015})}\BibitemShut {NoStop}%
\bibitem [{\citenamefont {Courtney}\ and\ \citenamefont
  {Bianconi}(2016)}]{Courtney:2016}%
  \BibitemOpen
  \bibfield  {author} {\bibinfo {author} {\bibfnamefont {O.~T.}\ \bibnamefont
  {Courtney}}\ and\ \bibinfo {author} {\bibfnamefont {G.}~\bibnamefont
  {Bianconi}},\ }\href {\doibase 10.1103/PhysRevE.93.062311} {\bibfield
  {journal} {\bibinfo  {journal} {Phys. Rev. E}\ }\textbf {\bibinfo {volume}
  {93}},\ \bibinfo {pages} {062311} (\bibinfo {year} {2016})}\BibitemShut
  {NoStop}%
\bibitem [{\citenamefont {Young}\ \emph {et~al.}(2017)\citenamefont {Young},
  \citenamefont {Petri}, \citenamefont {Vaccarino},\ and\ \citenamefont
  {Patania}}]{young2017construction}%
  \BibitemOpen
  \bibfield  {author} {\bibinfo {author} {\bibfnamefont {J.-G.}\ \bibnamefont
  {Young}}, \bibinfo {author} {\bibfnamefont {G.}~\bibnamefont {Petri}},
  \bibinfo {author} {\bibfnamefont {F.}~\bibnamefont {Vaccarino}}, \ and\
  \bibinfo {author} {\bibfnamefont {A.}~\bibnamefont {Patania}},\ }\href@noop
  {} {\bibfield  {journal} {\bibinfo  {journal} {Physical Review E}\ }\textbf
  {\bibinfo {volume} {96}},\ \bibinfo {pages} {032312} (\bibinfo {year}
  {2017})}\BibitemShut {NoStop}%
\bibitem [{\citenamefont {Benson}\ \emph {et~al.}(2018)\citenamefont {Benson},
  \citenamefont {Abebe}, \citenamefont {Schaub}, \citenamefont {Jadbabaie},\
  and\ \citenamefont {Kleinberg}}]{benson2018simplicial}%
  \BibitemOpen
  \bibfield  {author} {\bibinfo {author} {\bibfnamefont {A.~R.}\ \bibnamefont
  {Benson}}, \bibinfo {author} {\bibfnamefont {R.}~\bibnamefont {Abebe}},
  \bibinfo {author} {\bibfnamefont {M.~T.}\ \bibnamefont {Schaub}}, \bibinfo
  {author} {\bibfnamefont {A.}~\bibnamefont {Jadbabaie}}, \ and\ \bibinfo
  {author} {\bibfnamefont {J.}~\bibnamefont {Kleinberg}},\ }\href@noop {}
  {\bibfield  {journal} {\bibinfo  {journal} {arXiv preprint arXiv:1802.06916}\
  } (\bibinfo {year} {2018})}\BibitemShut {NoStop}%
\bibitem [{\citenamefont {Lord}\ \emph {et~al.}(2016)\citenamefont {Lord},
  \citenamefont {Expert}, \citenamefont {Fernandes}, \citenamefont {Petri},
  \citenamefont {Van~Hartevelt}, \citenamefont {Vaccarino}, \citenamefont
  {Deco}, \citenamefont {Turkheimer},\ and\ \citenamefont
  {Kringelbach}}]{lord2016insights}%
  \BibitemOpen
  \bibfield  {author} {\bibinfo {author} {\bibfnamefont {L.-D.}\ \bibnamefont
  {Lord}}, \bibinfo {author} {\bibfnamefont {P.}~\bibnamefont {Expert}},
  \bibinfo {author} {\bibfnamefont {H.~M.}\ \bibnamefont {Fernandes}}, \bibinfo
  {author} {\bibfnamefont {G.}~\bibnamefont {Petri}}, \bibinfo {author}
  {\bibfnamefont {T.~J.}\ \bibnamefont {Van~Hartevelt}}, \bibinfo {author}
  {\bibfnamefont {F.}~\bibnamefont {Vaccarino}}, \bibinfo {author}
  {\bibfnamefont {G.}~\bibnamefont {Deco}}, \bibinfo {author} {\bibfnamefont
  {F.}~\bibnamefont {Turkheimer}}, \ and\ \bibinfo {author} {\bibfnamefont
  {M.~L.}\ \bibnamefont {Kringelbach}},\ }\href@noop {} {\bibfield  {journal}
  {\bibinfo  {journal} {Frontiers in systems neuroscience}\ }\textbf {\bibinfo
  {volume} {10}},\ \bibinfo {pages} {85} (\bibinfo {year} {2016})}\BibitemShut
  {NoStop}%
\bibitem [{\citenamefont {Sizemore}\ \emph {et~al.}(2018)\citenamefont
  {Sizemore}, \citenamefont {Giusti}, \citenamefont {Kahn}, \citenamefont
  {Vettel}, \citenamefont {Betzel},\ and\ \citenamefont
  {Bassett}}]{sizemore2018cliques}%
  \BibitemOpen
  \bibfield  {author} {\bibinfo {author} {\bibfnamefont {A.~E.}\ \bibnamefont
  {Sizemore}}, \bibinfo {author} {\bibfnamefont {C.}~\bibnamefont {Giusti}},
  \bibinfo {author} {\bibfnamefont {A.}~\bibnamefont {Kahn}}, \bibinfo {author}
  {\bibfnamefont {J.~M.}\ \bibnamefont {Vettel}}, \bibinfo {author}
  {\bibfnamefont {R.~F.}\ \bibnamefont {Betzel}}, \ and\ \bibinfo {author}
  {\bibfnamefont {D.~S.}\ \bibnamefont {Bassett}},\ }\href@noop {} {\bibfield
  {journal} {\bibinfo  {journal} {Journal of computational neuroscience}\
  }\textbf {\bibinfo {volume} {44}},\ \bibinfo {pages} {115} (\bibinfo {year}
  {2018})}\BibitemShut {NoStop}%
\bibitem [{\citenamefont {Holme}\ and\ \citenamefont
  {Saram{\"a}ki}(2012)}]{Holme:2012}%
  \BibitemOpen
  \bibfield  {author} {\bibinfo {author} {\bibfnamefont {P.}~\bibnamefont
  {Holme}}\ and\ \bibinfo {author} {\bibfnamefont {J.}~\bibnamefont
  {Saram{\"a}ki}},\ }\href {\doibase
  http://dx.doi.org/10.1016/j.physrep.2012.03.001} {\bibfield  {journal}
  {\bibinfo  {journal} {Physics Reports}\ }\textbf {\bibinfo {volume} {519}},\
  \bibinfo {pages} {97 } (\bibinfo {year} {2012})}\BibitemShut {NoStop}%
\bibitem [{\citenamefont {Holme}(2015)}]{Holme:2015}%
  \BibitemOpen
  \bibfield  {author} {\bibinfo {author} {\bibfnamefont {P.}~\bibnamefont
  {Holme}},\ }\href {\doibase 10.1140/epjb/e2015-60657-4} {\bibfield  {journal}
  {\bibinfo  {journal} {The European Physical Journal B}\ }\textbf {\bibinfo
  {volume} {88}},\ \bibinfo {pages} {234} (\bibinfo {year} {2015})}\BibitemShut
  {NoStop}%
\bibitem [{\citenamefont {Stehl\'{e}}\ \emph {et~al.}(2010)\citenamefont
  {Stehl\'{e}}, \citenamefont {Barrat},\ and\ \citenamefont
  {Bianconi}}]{Stehle:2010}%
  \BibitemOpen
  \bibfield  {author} {\bibinfo {author} {\bibfnamefont {J.}~\bibnamefont
  {Stehl\'{e}}}, \bibinfo {author} {\bibfnamefont {A.}~\bibnamefont {Barrat}},
  \ and\ \bibinfo {author} {\bibfnamefont {G.}~\bibnamefont {Bianconi}},\
  }\href {\doibase 10.1103/PhysRevE.81.035101} {\bibfield  {journal} {\bibinfo
  {journal} {Physical Review E}\ }\textbf {\bibinfo {volume} {81}},\ \bibinfo
  {pages} {035101} (\bibinfo {year} {2010})}\BibitemShut {NoStop}%
\bibitem [{\citenamefont {Perra}\ \emph
  {et~al.}(2012{\natexlab{a}})\citenamefont {Perra}, \citenamefont
  {Gon\c{c}alves}, \citenamefont {Pastor-Satorras},\ and\ \citenamefont
  {Vespignani}}]{Perra:2012}%
  \BibitemOpen
  \bibfield  {author} {\bibinfo {author} {\bibfnamefont {N.}~\bibnamefont
  {Perra}}, \bibinfo {author} {\bibfnamefont {B.}~\bibnamefont
  {Gon\c{c}alves}}, \bibinfo {author} {\bibfnamefont {R.}~\bibnamefont
  {Pastor-Satorras}}, \ and\ \bibinfo {author} {\bibfnamefont {A.}~\bibnamefont
  {Vespignani}},\ }\href@noop {} {\bibfield  {journal} {\bibinfo  {journal}
  {Scientific reports}\ }\textbf {\bibinfo {volume} {2}},\ \bibinfo {pages}
  {469} (\bibinfo {year} {2012}{\natexlab{a}})}\BibitemShut {NoStop}%
\bibitem [{\citenamefont {Vestergaard}\ \emph {et~al.}(2014)\citenamefont
  {Vestergaard}, \citenamefont {G\'enois},\ and\ \citenamefont
  {Barrat}}]{Vestergaard:2014}%
  \BibitemOpen
  \bibfield  {author} {\bibinfo {author} {\bibfnamefont {C.~L.}\ \bibnamefont
  {Vestergaard}}, \bibinfo {author} {\bibfnamefont {M.}~\bibnamefont
  {G\'enois}}, \ and\ \bibinfo {author} {\bibfnamefont {A.}~\bibnamefont
  {Barrat}},\ }\href {\doibase 10.1103/PhysRevE.90.042805} {\bibfield
  {journal} {\bibinfo  {journal} {Phys. Rev. E}\ }\textbf {\bibinfo {volume}
  {90}},\ \bibinfo {pages} {042805} (\bibinfo {year} {2014})}\BibitemShut
  {NoStop}%
\bibitem [{\citenamefont {Karsai}\ \emph {et~al.}(2014)\citenamefont {Karsai},
  \citenamefont {Perra},\ and\ \citenamefont {Vespignani}}]{Karsai:2014}%
  \BibitemOpen
  \bibfield  {author} {\bibinfo {author} {\bibfnamefont {M.}~\bibnamefont
  {Karsai}}, \bibinfo {author} {\bibfnamefont {N.}~\bibnamefont {Perra}}, \
  and\ \bibinfo {author} {\bibfnamefont {A.}~\bibnamefont {Vespignani}},\
  }\href {\doibase 10.1038/srep04001} {\bibfield  {journal} {\bibinfo
  {journal} {Sci Rep}\ }\textbf {\bibinfo {volume} {4}},\ \bibinfo {pages}
  {4001} (\bibinfo {year} {2014})}\BibitemShut {NoStop}%
\bibitem [{\citenamefont {Sun}\ \emph {et~al.}(2015)\citenamefont {Sun},
  \citenamefont {Baronchelli},\ and\ \citenamefont {Perra}}]{Sun:2015}%
  \BibitemOpen
  \bibfield  {author} {\bibinfo {author} {\bibfnamefont {K.}~\bibnamefont
  {Sun}}, \bibinfo {author} {\bibfnamefont {A.}~\bibnamefont {Baronchelli}}, \
  and\ \bibinfo {author} {\bibfnamefont {N.}~\bibnamefont {Perra}},\
  }\href@noop {} {\bibfield  {journal} {\bibinfo  {journal} {Eur. Phys. J. B}\
  }\textbf {\bibinfo {volume} {88}},\ \bibinfo {pages} {326} (\bibinfo {year}
  {2015})}\BibitemShut {NoStop}%
\bibitem [{\citenamefont {Nadini}\ \emph {et~al.}(2018)\citenamefont {Nadini},
  \citenamefont {Sun}, \citenamefont {Ubaldi}, \citenamefont {Starnini},
  \citenamefont {Rizzo},\ and\ \citenamefont {Perra}}]{Nadini:2018}%
  \BibitemOpen
  \bibfield  {author} {\bibinfo {author} {\bibfnamefont {M.}~\bibnamefont
  {Nadini}}, \bibinfo {author} {\bibfnamefont {K.}~\bibnamefont {Sun}},
  \bibinfo {author} {\bibfnamefont {E.}~\bibnamefont {Ubaldi}}, \bibinfo
  {author} {\bibfnamefont {M.}~\bibnamefont {Starnini}}, \bibinfo {author}
  {\bibfnamefont {A.}~\bibnamefont {Rizzo}}, \ and\ \bibinfo {author}
  {\bibfnamefont {N.}~\bibnamefont {Perra}},\ }\href {\doibase
  10.1038/s41598-018-20908-x} {\bibfield  {journal} {\bibinfo  {journal}
  {Scientific Reports}\ }\textbf {\bibinfo {volume} {8}},\ \bibinfo {pages}
  {2352} (\bibinfo {year} {2018})}\BibitemShut {NoStop}%
\bibitem [{\citenamefont {Kim}\ \emph {et~al.}(2018)\citenamefont {Kim},
  \citenamefont {Ha},\ and\ \citenamefont {Jeong}}]{Kim:2018}%
  \BibitemOpen
  \bibfield  {author} {\bibinfo {author} {\bibfnamefont {H.}~\bibnamefont
  {Kim}}, \bibinfo {author} {\bibfnamefont {M.}~\bibnamefont {Ha}}, \ and\
  \bibinfo {author} {\bibfnamefont {H.}~\bibnamefont {Jeong}},\ }\href
  {\doibase 10.1103/PhysRevE.97.062148} {\bibfield  {journal} {\bibinfo
  {journal} {Phys. Rev. E}\ }\textbf {\bibinfo {volume} {97}},\ \bibinfo
  {pages} {062148} (\bibinfo {year} {2018})}\BibitemShut {NoStop}%
\bibitem [{\citenamefont {Perra}\ \emph
  {et~al.}(2012{\natexlab{b}})\citenamefont {Perra}, \citenamefont
  {Baronchelli}, \citenamefont {Mocanu}, \citenamefont
  {Gon\ifmmode~\mbox{\c{c}}\else \c{c}\fi{}alves}, \citenamefont
  {Pastor-Satorras},\ and\ \citenamefont {Vespignani}}]{Perra:2012b}%
  \BibitemOpen
  \bibfield  {author} {\bibinfo {author} {\bibfnamefont {N.}~\bibnamefont
  {Perra}}, \bibinfo {author} {\bibfnamefont {A.}~\bibnamefont {Baronchelli}},
  \bibinfo {author} {\bibfnamefont {D.}~\bibnamefont {Mocanu}}, \bibinfo
  {author} {\bibfnamefont {B.}~\bibnamefont {Gon\ifmmode~\mbox{\c{c}}\else
  \c{c}\fi{}alves}}, \bibinfo {author} {\bibfnamefont {R.}~\bibnamefont
  {Pastor-Satorras}}, \ and\ \bibinfo {author} {\bibfnamefont {A.}~\bibnamefont
  {Vespignani}},\ }\href {\doibase 10.1103/PhysRevLett.109.238701} {\bibfield
  {journal} {\bibinfo  {journal} {Phys. Rev. Lett.}\ }\textbf {\bibinfo
  {volume} {109}},\ \bibinfo {pages} {238701} (\bibinfo {year}
  {2012}{\natexlab{b}})}\BibitemShut {NoStop}%
\bibitem [{\citenamefont {Barrat}\ \emph
  {et~al.}(2008{\natexlab{b}})\citenamefont {Barrat}, \citenamefont
  {Barthelemy},\ and\ \citenamefont {Vespignani}}]{barrat2008dynamical}%
  \BibitemOpen
  \bibfield  {author} {\bibinfo {author} {\bibfnamefont {A.}~\bibnamefont
  {Barrat}}, \bibinfo {author} {\bibfnamefont {M.}~\bibnamefont {Barthelemy}},
  \ and\ \bibinfo {author} {\bibfnamefont {A.}~\bibnamefont {Vespignani}},\
  }\href@noop {} {\emph {\bibinfo {title} {Dynamical processes on complex
  networks}}}\ (\bibinfo  {publisher} {Cambridge university press},\ \bibinfo
  {year} {2008})\BibitemShut {NoStop}%
\bibitem [{\citenamefont {Hatcher}(2005)}]{hatcher2005algebraic}%
  \BibitemOpen
  \bibfield  {author} {\bibinfo {author} {\bibfnamefont {A.}~\bibnamefont
  {Hatcher}},\ }\href@noop {} {\emph {\bibinfo {title} {Algebraic topology}}}\
  (\bibinfo {year} {2005})\BibitemShut {NoStop}%
\bibitem [{\citenamefont {Palla}\ \emph {et~al.}(2005)\citenamefont {Palla},
  \citenamefont {Der{\'e}nyi}, \citenamefont {Farkas},\ and\ \citenamefont
  {Vicsek}}]{palla2005uncovering}%
  \BibitemOpen
  \bibfield  {author} {\bibinfo {author} {\bibfnamefont {G.}~\bibnamefont
  {Palla}}, \bibinfo {author} {\bibfnamefont {I.}~\bibnamefont {Der{\'e}nyi}},
  \bibinfo {author} {\bibfnamefont {I.}~\bibnamefont {Farkas}}, \ and\ \bibinfo
  {author} {\bibfnamefont {T.}~\bibnamefont {Vicsek}},\ }\href@noop {}
  {\bibfield  {journal} {\bibinfo  {journal} {Nature}\ }\textbf {\bibinfo
  {volume} {435}},\ \bibinfo {pages} {814} (\bibinfo {year}
  {2005})}\BibitemShut {NoStop}%
\bibitem [{\citenamefont {Horak}\ and\ \citenamefont
  {Jost}(2013)}]{horak2013spectra}%
  \BibitemOpen
  \bibfield  {author} {\bibinfo {author} {\bibfnamefont {D.}~\bibnamefont
  {Horak}}\ and\ \bibinfo {author} {\bibfnamefont {J.}~\bibnamefont {Jost}},\
  }\href@noop {} {\bibfield  {journal} {\bibinfo  {journal} {Advances in
  Mathematics}\ }\textbf {\bibinfo {volume} {244}},\ \bibinfo {pages} {303}
  (\bibinfo {year} {2013})}\BibitemShut {NoStop}%
\bibitem [{\citenamefont {Mukherjee}\ and\ \citenamefont
  {Steenbergen}(2016)}]{mukherjee2016random}%
  \BibitemOpen
  \bibfield  {author} {\bibinfo {author} {\bibfnamefont {S.}~\bibnamefont
  {Mukherjee}}\ and\ \bibinfo {author} {\bibfnamefont {J.}~\bibnamefont
  {Steenbergen}},\ }\href@noop {} {\bibfield  {journal} {\bibinfo  {journal}
  {Random structures \& algorithms}\ }\textbf {\bibinfo {volume} {49}},\
  \bibinfo {pages} {379} (\bibinfo {year} {2016})}\BibitemShut {NoStop}%
\bibitem [{\citenamefont {Parzanchevski}\ and\ \citenamefont
  {Rosenthal}(2017)}]{parzanchevski2017simplicial}%
  \BibitemOpen
  \bibfield  {author} {\bibinfo {author} {\bibfnamefont {O.}~\bibnamefont
  {Parzanchevski}}\ and\ \bibinfo {author} {\bibfnamefont {R.}~\bibnamefont
  {Rosenthal}},\ }\href@noop {} {\bibfield  {journal} {\bibinfo  {journal}
  {Random Structures \& Algorithms}\ }\textbf {\bibinfo {volume} {50}},\
  \bibinfo {pages} {225} (\bibinfo {year} {2017})}\BibitemShut {NoStop}%
\bibitem [{\citenamefont {Muhammad}\ and\ \citenamefont
  {Egerstedt}(2006)}]{muhammad2006control}%
  \BibitemOpen
  \bibfield  {author} {\bibinfo {author} {\bibfnamefont {A.}~\bibnamefont
  {Muhammad}}\ and\ \bibinfo {author} {\bibfnamefont {M.}~\bibnamefont
  {Egerstedt}},\ }in\ \href@noop {} {\emph {\bibinfo {booktitle} {Proc. of 17th
  International Symposium on Mathematical Theory of Networks and Systems}}}\
  (\bibinfo {organization} {Citeseer},\ \bibinfo {year} {2006})\ pp.\ \bibinfo
  {pages} {1024--1038}\BibitemShut {NoStop}%
\bibitem [{\citenamefont {Friedman}(1998)}]{friedman1998computing}%
  \BibitemOpen
  \bibfield  {author} {\bibinfo {author} {\bibfnamefont {J.}~\bibnamefont
  {Friedman}},\ }\href@noop {} {\bibfield  {journal} {\bibinfo  {journal}
  {Algorithmica}\ }\textbf {\bibinfo {volume} {21}},\ \bibinfo {pages} {331}
  (\bibinfo {year} {1998})}\BibitemShut {NoStop}%
\bibitem [{Soc()}]{SocioPatterns}%
  \BibitemOpen
  \href@noop {} {\enquote {\bibinfo {title}
  {{\url{http://www.sociopatterns.org/}}},}\ }\BibitemShut {NoStop}%
\bibitem [{\citenamefont {Watts}(2002)}]{Watts:2002}%
  \BibitemOpen
  \bibfield  {author} {\bibinfo {author} {\bibfnamefont {D.~J.}\ \bibnamefont
  {Watts}},\ }\href {\doibase 10.1073/pnas.082090499} {\bibfield  {journal}
  {\bibinfo  {journal} {Proceedings of the National Academy of Sciences}\
  }\textbf {\bibinfo {volume} {99}},\ \bibinfo {pages} {5766} (\bibinfo {year}
  {2002})},\ \Eprint
  {http://arxiv.org/abs/http://www.pnas.org/content/99/9/5766.full.pdf}
  {http://www.pnas.org/content/99/9/5766.full.pdf} \BibitemShut {NoStop}%
\bibitem [{\citenamefont {Karimi}\ and\ \citenamefont
  {Holme}(2013)}]{Karimi:2013}%
  \BibitemOpen
  \bibfield  {author} {\bibinfo {author} {\bibfnamefont {F.}~\bibnamefont
  {Karimi}}\ and\ \bibinfo {author} {\bibfnamefont {P.}~\bibnamefont {Holme}},\
  }\href {\doibase https://doi.org/10.1016/j.physa.2013.03.050} {\bibfield
  {journal} {\bibinfo  {journal} {Physica A: Statistical Mechanics and its
  Applications}\ }\textbf {\bibinfo {volume} {392}},\ \bibinfo {pages} {3476 }
  (\bibinfo {year} {2013})}\BibitemShut {NoStop}%
\end{thebibliography}
\end{document}